 \documentclass[manuscript=article]{achemso}
 \usepackage{natbib}

\usepackage[version=3]{mhchem} 
\usepackage{subfig} %
\usepackage[dvipsnames]{xcolor}

\author{Bruno Ipaves}
\affiliation{Instituto de F\'{\i}sica, Universidade de S\~ao Paulo, \\
CP 66318, CEP 05315-970 S\~ao Paulo - SP, Brazil}
\author{Jo\~ao F. Justo}
\affiliation{Escola Polit\'ecnica, Universidade de S\~ao Paulo, \\
CEP 05508-970, S\~ao Paulo - SP, Brazil}
\author{Lucy V. C. Assali}
\affiliation{Instituto de F\'{\i}sica, Universidade de S\~ao Paulo, \\
CP 66318, CEP 05315-970, S\~ao Paulo - SP, Brazil}
\email{ipaves@if.usp.br, jjusto@lme.usp.br, lassali@if.usp.br}

\title[An \textsf{achemso} demo]
{Carbon-related Bilayers: 
Nanoscale Building Blocks for Self-Assembly Nanomanufacturing}

\begin{document}

\begin{tocentry}





\includegraphics[scale = 1, trim={0cm 0.0cm 0cm 0cm}, clip]{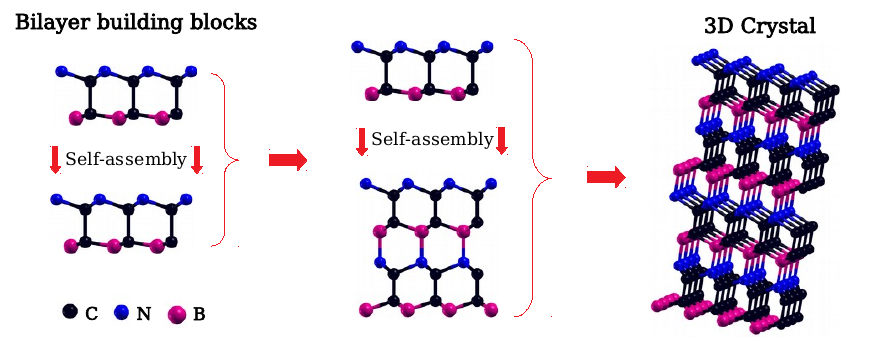}

\end{tocentry}
\begin{abstract}

Using a first-principles total energy methodology, we investigated the properties of graphene-like carbon mono and bilayers, functionalized with nitrogen and boron atoms. The resulting stable structures were explored in terms of their potential use as nanoscale two-dimensional building blocks for self-assembly of macroscopic structures. We initially considered graphene monolayers functionalized with nitrogen and boron, but none of them was dynamically stable, in terms of the respective layer phonon spectra. Then, we considered the functionalized graphene-like bilayers (labeled as \ce{NCCN}, \ce{NCNC}, \ce{BCCB}, and NCCB), analyzing their stability, electronic and mechanical properties, and chemical reactivity. We found that while the \ce{NCCN}, \ce{NCNC}, and NCCB bilayers were stable, the BCCB one was not. Additionally, the NCCN and NCCB bilayers were explored as potential two-dimensional building blocks for nanostructure self-assembly, which could form stable bulk structures. Particularly, the NCCB bilayer seemed the best choice as a building block, since the resulting 3D crystals, formed by stacking NCCB bilayers, were energetically stable.

\end{abstract}

\section{Introduction}

Over the last thirty years, a myriad of nanostructures has been identified, which led to the synthesis of an even greater number of nanostructures.\cite{natelson} The possibility of controlling and manipulating matter at the nanoscopic level, either by artificial or natural processes, has been very appealing, representing a major revolution in science and technology in many aspects. This ongoing revolution has brought the opportunity of building nanostructures with tailored physical and chemical properties or functionalities. Accordingly, it is easily foreseeable the potential applications for these nanostructures in the most diverse areas, such as electronics, photovoltaics, energy storage, quantum computing, and medicine.\cite{Bhushan,Huang2011}   

Many of those nanostructures could be used as nanoscale building blocks, also labeled as molecular building blocks \cite{Merkle}, to function as nanobricks to assemble complex structures at mesoscopic or macroscopic levels.\cite{Gooding, Garcia} Nevertheless, one of the main challenges has been associated with the difficulty in manipulating those nanostructures. Particularly, many techniques allow manipulation of only a single nanostructure at a time, which is unrealistically too slow to build structures at a large scale.\cite{lu2007}  Alternatively, the interatomic or intermolecular interactions could drive the assembly processes, which has been labeled as self-assembly.

A self-assembly process generally uses a set of nanostructures as building blocks, initially in a disordered configuration, which is ordered in a tailored macroscopic configuration, taking the advantage of the local interactions between those blocks, without the influence of external agents.\cite{Liu} This process typically evolves based on the weak interactions between the building blocks, such as van der Waals or hydrogen bonds, rather than with stronger ones, such as ionic or covalent interactions. This process, based on a bottom-up technique, seems very promising over the next few years, in contrast to traditional and widely used top-down processes.\cite{Nagarajan} 

The key ingredient for self-assembly is to use an appropriate building block. Zero-dimensional nanostructures, such as large molecules \cite{Merkle,grzelczak2010directed}, and one-dimensional nanostructures, such as nanowires and nanotubes \cite{Fan}, have been widely explored to serve as building blocks. On the other hand, so far only very few investigations have considered two-dimensional (2D) nanostructures, such as nanosheets, as building blocks \cite{Sun}. Considering the recent rising of graphene, along with a vast amount of other two-dimensional sheets that have been recently synthesized \cite{miro2014atlas}, it is interesting to explore configurations, architectures, and functionalizations that those nanosheets could provide to serve as building blocks.

Within such context, it is important to explore the electronic and structural properties of potential 2D nanostructures that could serve as nanobricks for self-assembly processes. Over the last few years, a vast number of 2D materials have been identified or synthesized \cite{garcia2011group, miro2014atlas}. Those materials present considerably different physical properties, while graphene is a semimetal with a zero bandgap, \ce{MoS2} sheet is a semiconductor, and h-BN is an insulator \cite{akinwande2017review}. Moreover, functionalizing those structures modify radically their properties, for example, graphane, the hydrogenated counterpart of graphene, is an insulator \cite{garcia2011group}. The specific properties of each 2D nanostructure determines if it is appropriate for self-assembly processes.

 Heterostructures have been assembled by stacking atomic monolayers, which remained stable and bounded to each other by the van der Waals (vdW) interactions between neighboring layers.\cite{fang2014,novoselov20162d} Although those out-of-plane interactions are weak, when compared to the in-plane covalent interactions, they are strong enough to keep the stacks together.\cite{geim2013van} Several heterostructures based on graphene, such as graphene monolayers sandwiched in hexagonal boron nitride, have been studied by experimental and theoretical investigations.\cite{britnell2012field, haigh2012cross, ma2014tunable} However, despite the potential applications, the resulting heterostructures are generally too weak, due to the type of interplanar interactions, which could bring issues on their stability and robustness. Moreover, there is no strong driving force for self-assembly beyond the van der Waals interactions. Additionally, there is a major challenge in manipulating the monolayers using typical techniques, such as the atomic force microscopy, which essentially builds a layer-by-layer heterostructure. Therefore, there is an opportunity to establish new methodologies to assemble heterostructures out of 2D building blocks, in order to optimize the assembling procedure and the physical properties of the resulting structures. 

Here, we used first-principles calculations to explore the properties of carbon-related functionalized monolayers and bilayers, and the possibility of their use as two-dimensional building blocks for self-assembly nanomanufacturing, the same way as for other types of building blocks.\cite{Merkle,Fan} Since graphene and its multi-layer graphene structures provide out-of-plane weak interactions, they would not be the leading candidates to serve as building blocks. Therefore, we searched for other stable two-dimensional structures that could serve as building blocks. The initial attempt was functionalizing a graphene monolayer, with nitrogen and boron atoms, but the resulting structures were mechanically unstable. Then, we explored the structural properties and stability of graphene-like bilayers. In that sense, we investigated the properties of substitutional nitrogen and/or boron atoms in graphene bilayers, in several configurations, and found that many of them were stable. We also explored the interactions of those stable bilayers with other bilayers and found a number of possible structures that could serve as two-dimensional building blocks. Particularly, the interactions between neighboring bilayers could be strong enough to provide stable 3D bulk structures with several desired properties. Those strong structures would contrast with typical self-assembled ones, which allow reversible disassemble easily and are unstable under even mild temperatures. 


\section{Computational  Details}

The calculations were performed using the Quantum ESPRES\-SO computational package \cite{Giannozzi2009}. The electronic interactions were described within the density functional theory (DFT) framework, in which the exchange-correlation potential included the van der Waals interactions, within the Dion {\it et al.} scheme \cite{dion2004} and optimized by Klimes {\it et al.} (optB88) \cite{klimevs2009}. The electronic wave-functions were described by a projector augmented-wave method (PAW) \cite{Kresse}, taking a plane-wave basis set with an energy cutoff of 1100 eV. 

For all calculations, convergence in total energy was set to 0.1 meV/atom between two self-consistent iteractions. Structural optimization was performed by considering relaxation in all ions, without symmetry constraints, until forces were lower than 1 meV/{\AA} in any ion. The Brillouin zones for computing the electronic states were sampled by a $16 \times 16 \times 1$ Monkhorst-Pack $k$-point grid \cite{Monkhorst}. 

The 2D structures (monolayers and bilayers) were built using periodic boundary conditions with a hexagonal simulation cell. In the perpendicular direction to the sheets ($z$ axis), a lattice parameter of at least 15 {\AA} was used, which was large enough to prevent interactions with the cell images in that direction. The vibrational normal modes (phonons) of the systems were computed by the phonon dispersion dynamic matrix, using the density functional perturbation theory \cite{baroni2001phonons}. The Brillouin zones were sampled by an $8 \times 8 \times 2$ $q$-point mesh in order to obtain the phonon dispersion properties. Such theoretical framework and convergence criteria have been shown to provide a reliable description of a number of carbon-related nanosystems \cite{klimevs2011}.

In order to confirm the validity of all approximations used in this investigation, we compared our results on the electronic and vibrational properties of graphene and graphite with available experimental and theoretical data. For both systems, the lattice parameter and the carbon-carbon interatomic distance were 2.47 {\AA} and 1.42 {\AA}, respectively, and the graphite $c$ lattice parameter was 6.68 {\AA}. These results were in excellent agreement with the respective experimental and theoretical values reported in the literature \cite{trucano,Reich2002,neto2009electronic}. Additionally, the electronic and vibrational properties were well described, showing that this methodology provided an appropriate description of the systems of interest here.  

Total energy minimizations, with respect to the changes in the interatomic distance $h$ between layers, were performed by relaxing the positions of all atoms of the system. For example, the calculation of the relative total energy as a function of the interplanar carbon-carbon distance $h_{\rm C-C}$ for the $AB$-NCCN bilayer led to a minimum energy with $h_{\rm C-C} = 1.576$ {\AA}. The same procedure was applied to all of the stable bilayers, in order to find the equilibrium interplanar distance between two monolayers.


The energy of formation $E_f$ of any bilayer was computed by
\begin{equation}
 E_f = E_{\rm tot} ({\rm C}_2{\rm N}_x{\rm B}_y) - 2E ({\rm C}) - xE({\rm N}) -yE({\rm B}),
 \label{eq1}
 \end{equation}
where $E_{\rm tot} ({\rm C}_2{\rm N}_x{\rm B}_y)$ is the total energy of the bilayer, per unit formula, with two carbon, $x$ nitrogen, and $y$ boron atoms. The $E ({\rm C})$, $E({\rm N})$, and $E({\rm B})$ are the total energies (per atom) of respectively carbon, nitrogen, and boron, in their standard reference states. Those energies, computed within the same methodology described earlier, were obtained from the total energy of carbon in the graphene lattice, nitrogen in an isolated N$_2$ molecule, and boron in a trigonal crystalline structure ($\beta$-boron). The total energy of the diatomic molecule was obtained considering a large three-dimensional simulation cell. This methodology to compute the energy of formation has been used in several other investigations in the literature \cite{ayres2006role,larico2009electronic}. 

The bilayer-crystal binding (exfoliation) energy $E_b$ was computed by
\begin{equation}
 E_b = E_{\rm tot} (\text{crystal}) - E_{\rm tot} (\text{building-block}),
 \label{eq1}
 \end{equation}
where $E_{\rm tot} ({\rm crystal})$ is the total energy of a hexagonal crystalline structure, where the basis consist of one or two bilayers, and $E_{\rm tot} (\text{building-block})$ is the total energy of the bilayer unit formula. By using this methodology, the value acquired for graphite was $E_b = E_{\rm tot} (\text{graphite}) - E_{\rm tot} (\text{graphene}) =$ 70 meV/atom, which was in good agreement with the reported theoretical and experimental values from other investigations \cite{lebedeva2011interlayer}.

\section{Results}

\begin{figure}[htb]
\centering
\subfloat[]{
\includegraphics[scale = 0.15, trim={3.5cm 2.5cm 0cm 3.35cm}, clip]{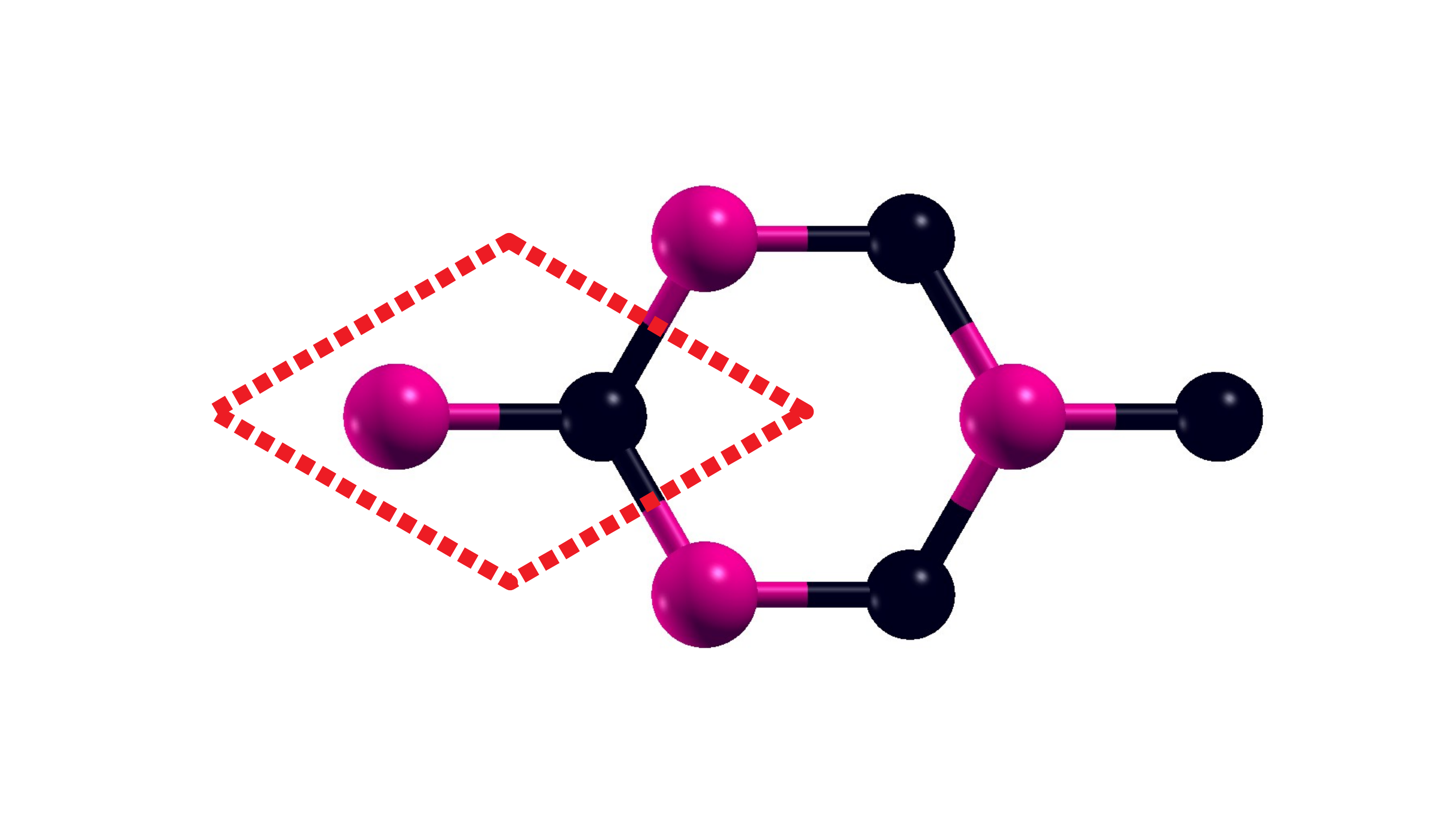} \quad
\includegraphics[scale = 0.15, trim={3.5cm 2.5cm 0cm 3.35cm}, clip]{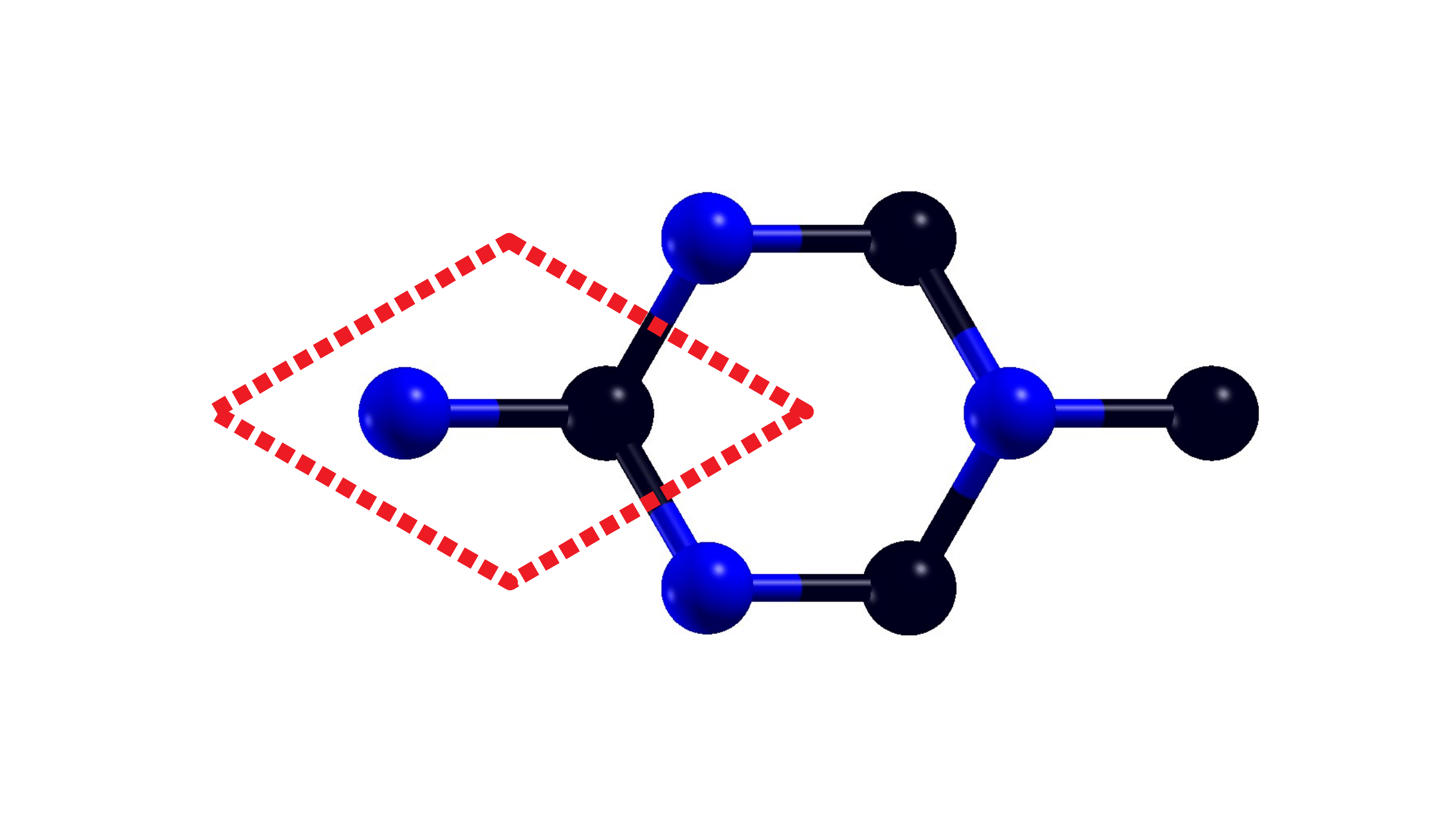}} \quad
\subfloat[]{
\includegraphics[scale = 0.15, trim={0cm 0.0cm 0cm 0cm}, clip]{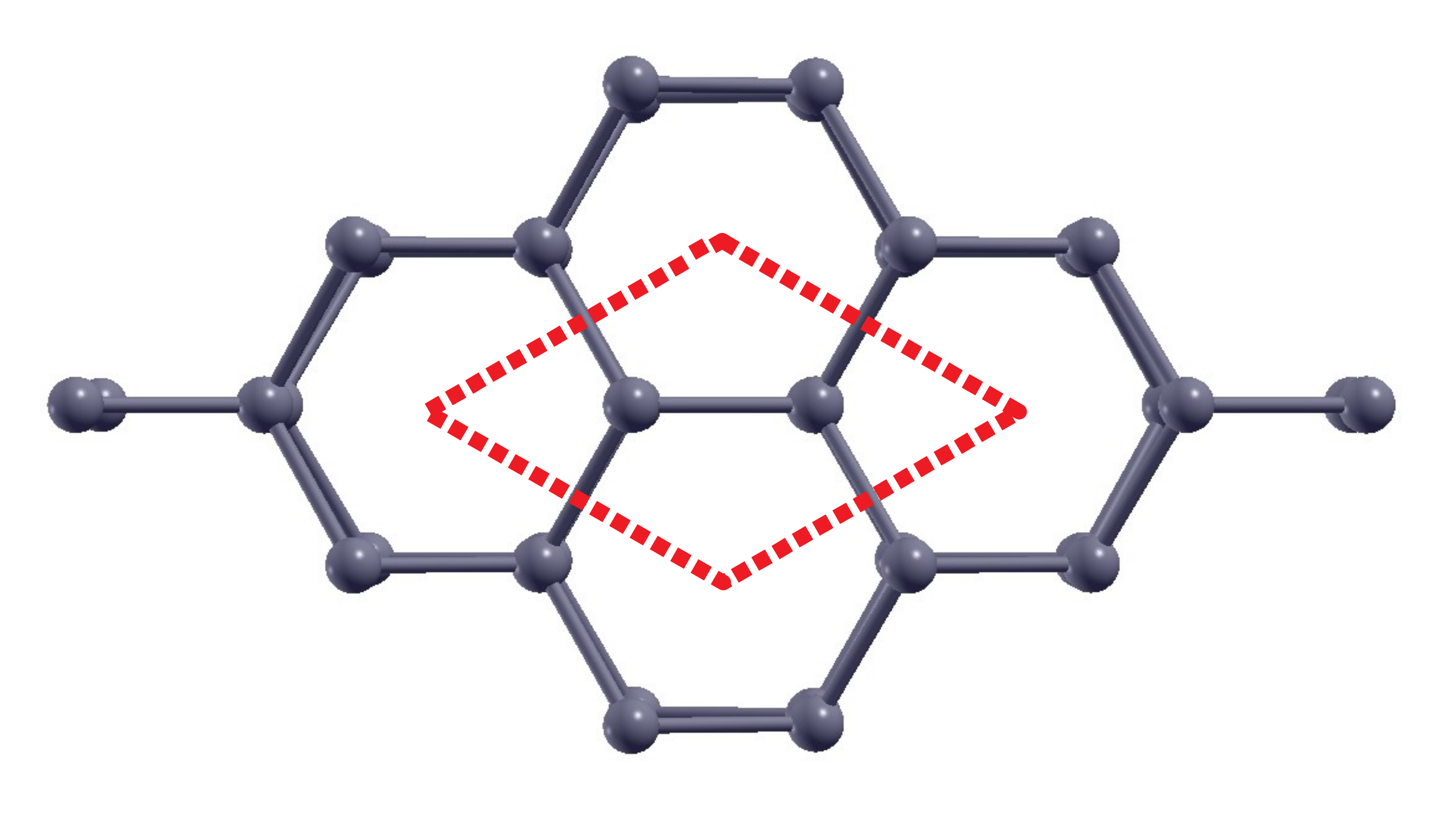}   \quad
\includegraphics[scale = 0.15, trim={0cm 0.0cm 0cm 0cm}, clip]{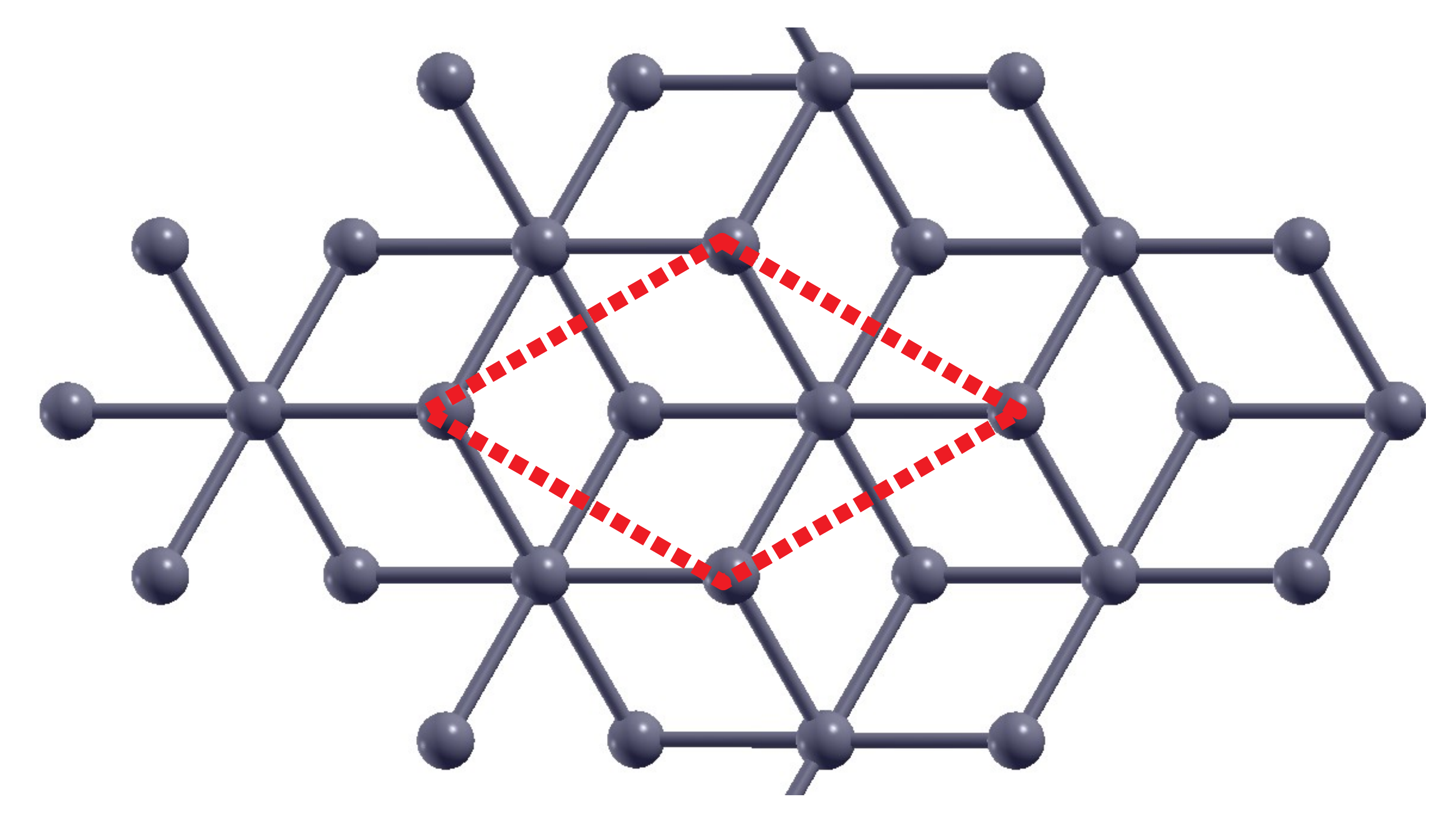} \quad
\includegraphics[scale = 0.15, trim={0cm 0.0cm 0cm 0cm}, clip]{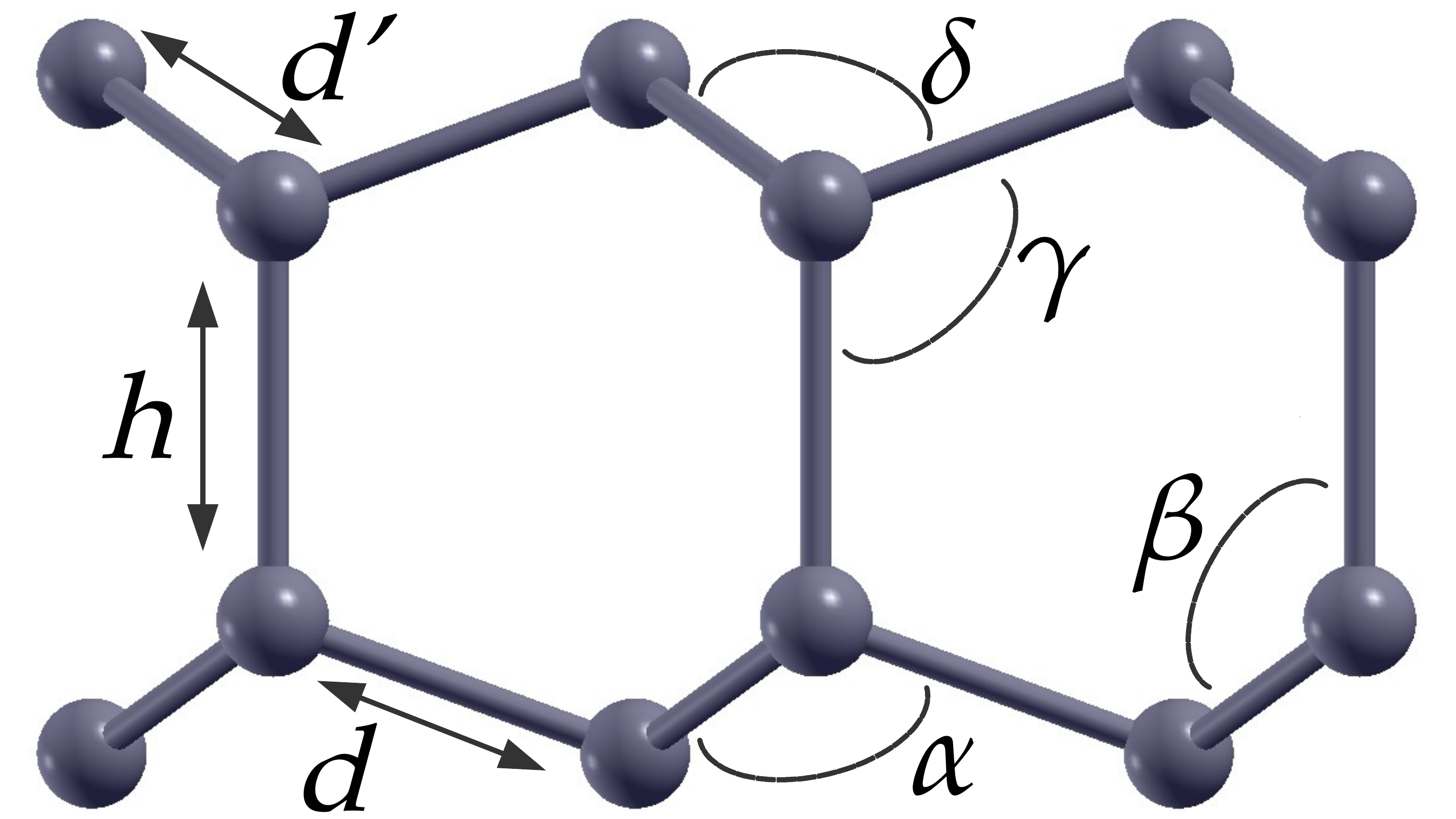}} \quad
\subfloat[]{
\includegraphics[scale = 0.5, trim={0cm 5.0cm 0cm 5cm}, clip]{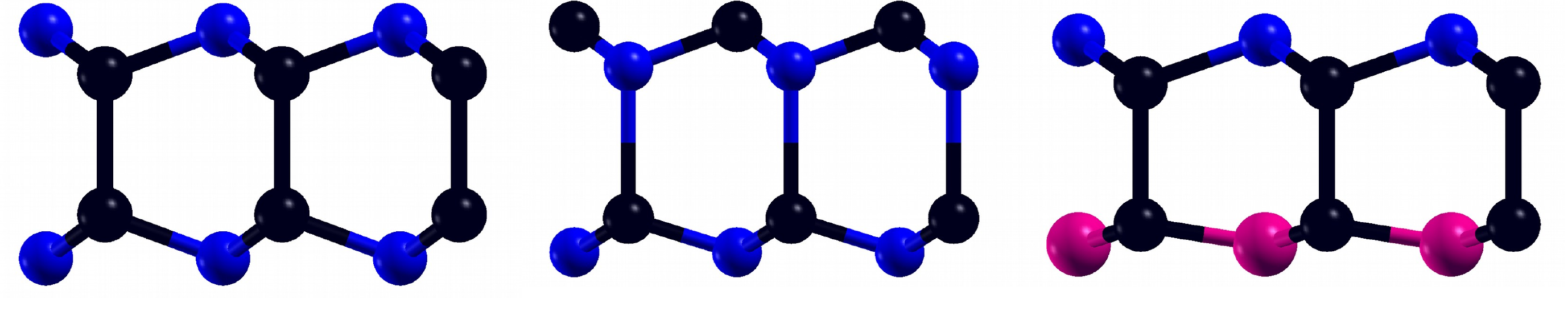}}\quad
\caption{Schematic representation of (a) the top view of the monolayers 50\% doped graphene with B or N atoms; (b) the top view of a bilayer in the $AA$-stacking (left) and $AB$-stacking (middle), and the side view of a bilayer (right); (c) the NCCN (left), the NCNC (middle) and the NCCB (rigth) bilayers investigated here. The black, pink, and blue spheres represent carbon, boron, and nitrogen atoms, respectively. The red dashed lines represent the limits of the unit cells in all structures. The right side of panel (b) shows the labels given for the inter-layer distance $h$, the interatomic distances $d$, and $d^{\prime}$ between atoms in each layer of the system. $\alpha, \beta, \gamma$, and $\delta$ are the bond angles.}
\label{AA_AB}
\end{figure}

\textbf{Stability of monolayers and bilayers}.
Figure \ref{AA_AB} shows a schematic representation of the configurations explored here. The first structure investigated consisted of a graphene sheet 50\% doped with substitutional nitrogen (h-CN) or boron (h-CB) atoms, as shown in figure \ref{AA_AB} (a). Since the atomic sizes of nitrogen and boron atoms are similar to the one of carbon, the h-CN and the h-CB monolayer systems presented small relaxations with respect to the original graphene, remaining in the hexagonal symmetry and in a flat configuration. However, they were dynamically unstable, presenting phonon spectra with imaginary frequencies in some branches, consistent with results from other theoretical investigations \cite{shi2015,zhou2015}. Therefore, in order to overcome the dynamic instability of these monolayers, and still obtain an ordered and stable 50\% N- or B-doped graphene-like systems, we explored the stability and properties of graphene-like bilayers, with the $AA$- and $AB$-stacking structures, doped with N and/or B atoms. These structures are displayed in figures \ref{AA_AB} (b) and (c), along with the labels given for the interatomic distances and bond angles. 

A large number of bilayers were initially considered in this investigation, but only six structures were dynamically stable. A previous investigation suggested that two-dimensio\-nal hexagonal lattice of carbon nitride structures should be unstable with N concentration exceeding 37.5\% \cite{shi2015}. However, we found 
stable structures presenting a 1:1 stoichiometry, formed by stacking two h-CN monolayers, i.e., forming a 50\% N-doped graphene-like bilayer. According to the schematic atomic structure representations shown in figure \ref{AA_AB}, the stable 50\% N-doped graphene-like bilayers were labeled as AA-NCCN, AB-NCCN, AA-NCNC, and AB-NCNC. The other two systems that were also dynamically stable were formed by stacking an h-CN layer with an h-CB one, i.e., a 25\% N- and 25\% B-doped graphene-like bilayer. These bilayers were labeled as $AA$-NCCB and $AB$-NCCB. We also explored the bilayers formed by stacking two h-CB layers, i.e., the 50\% B-doped graphene-like bilayers (BCCB), and found no dynamically stable structure. 

\begin{figure}[h!]
 \centering
 \subfloat[]{
 \includegraphics[scale = 0.45, trim={0cm 0.0cm 0cm 0.0cm}, clip]{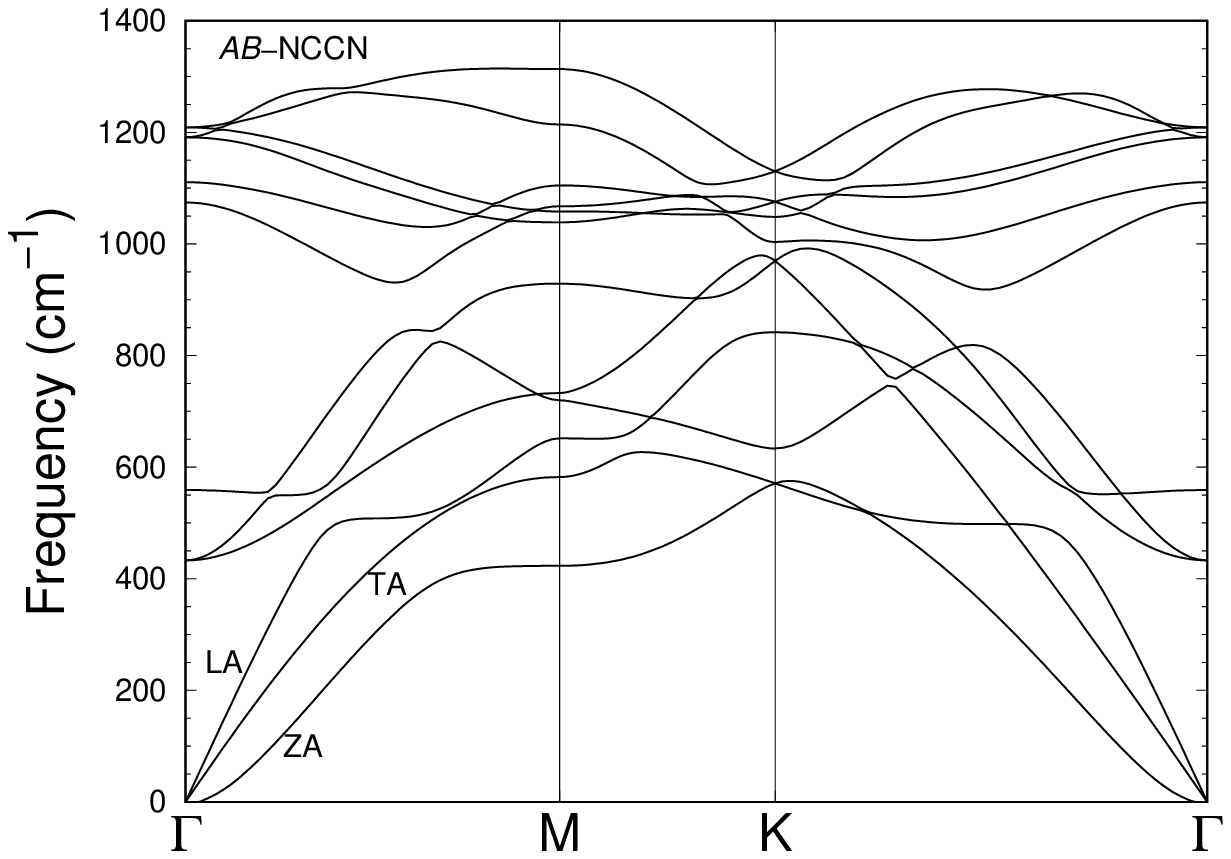}} 
 \subfloat[]{
 \includegraphics[scale = 0.45, trim={0cm 0.0cm 0cm 0.0cm}, clip]{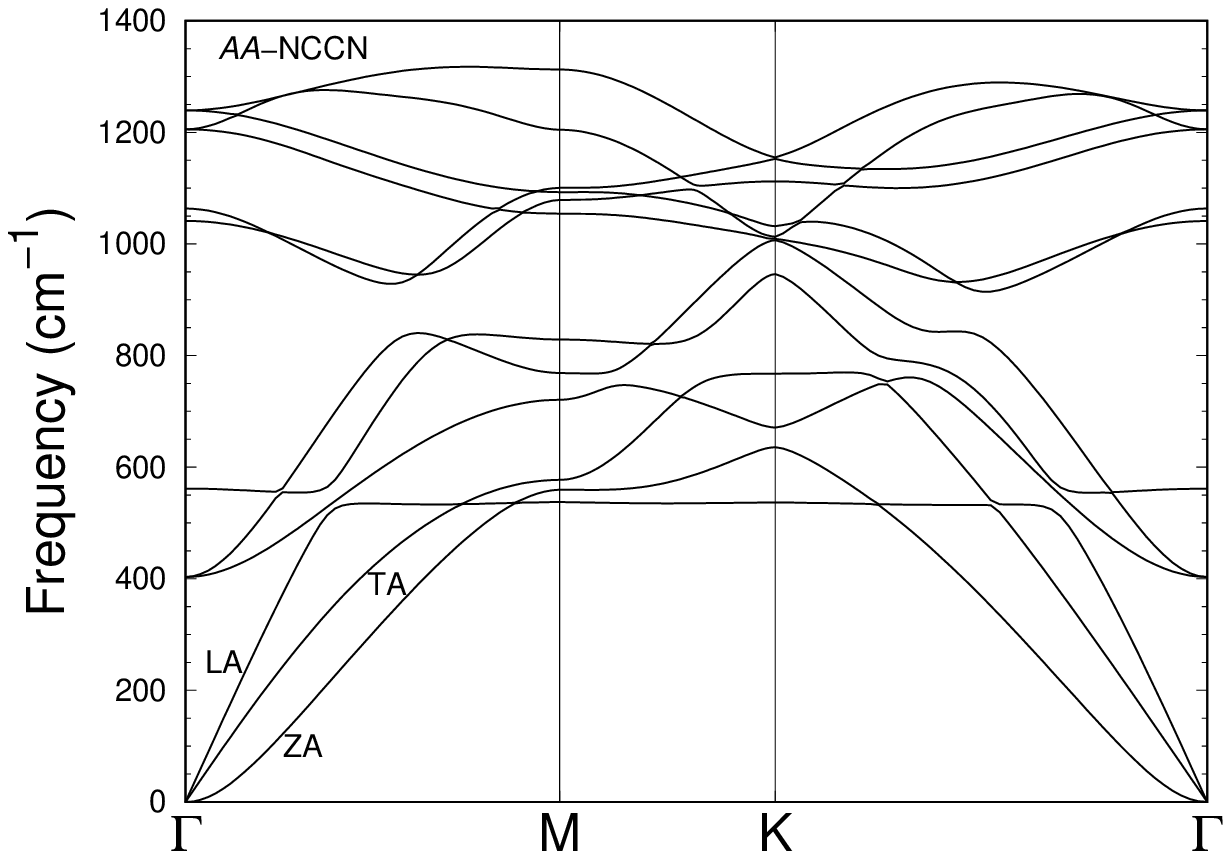}} \quad
 \subfloat[]{
 \includegraphics[scale = 0.45, trim={0cm 0.0cm 0cm 0.0cm}, clip]{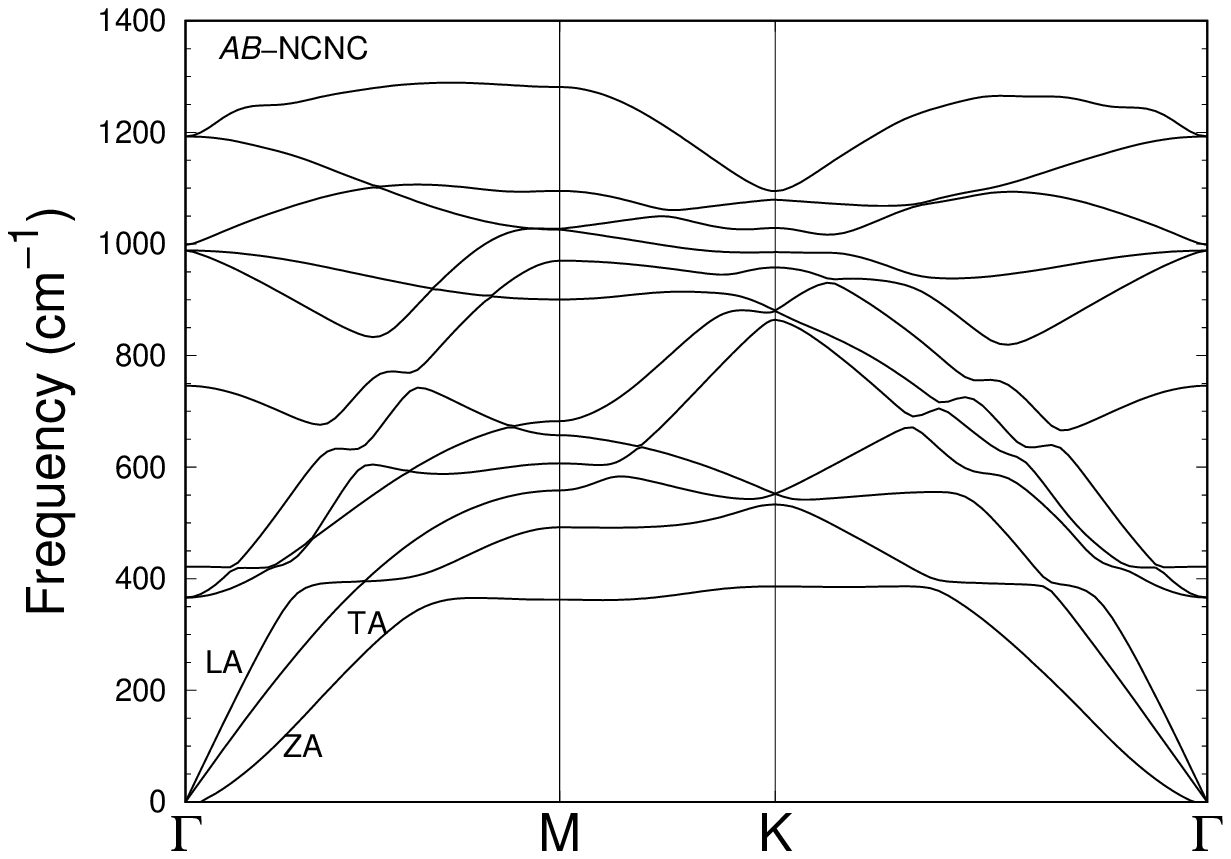}} 
 \subfloat[]{
 \includegraphics[scale = 0.45, trim={0cm 0.0cm 0cm 0.0cm}, clip]{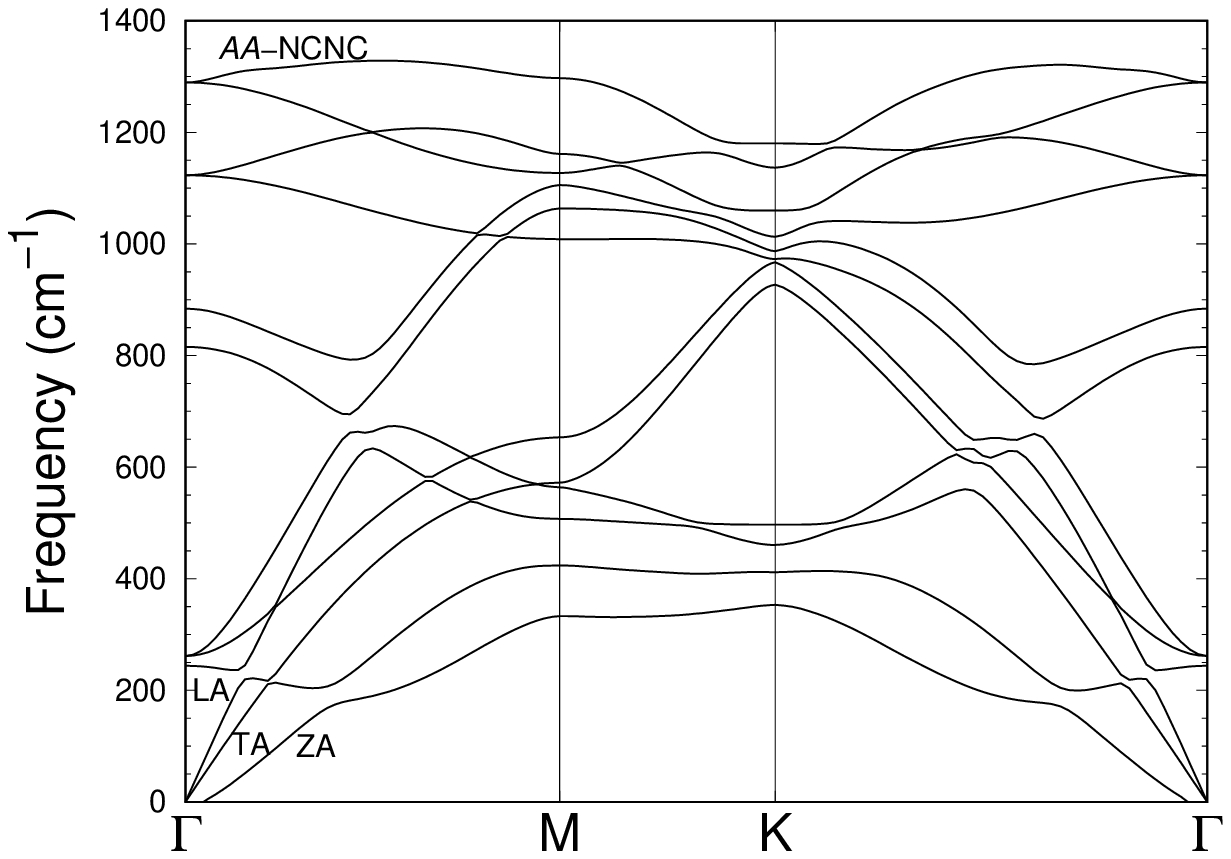}} \quad
 \subfloat[]{
 \includegraphics[scale = 0.45, trim={0cm 0.0cm 0cm 0.0cm}, clip]{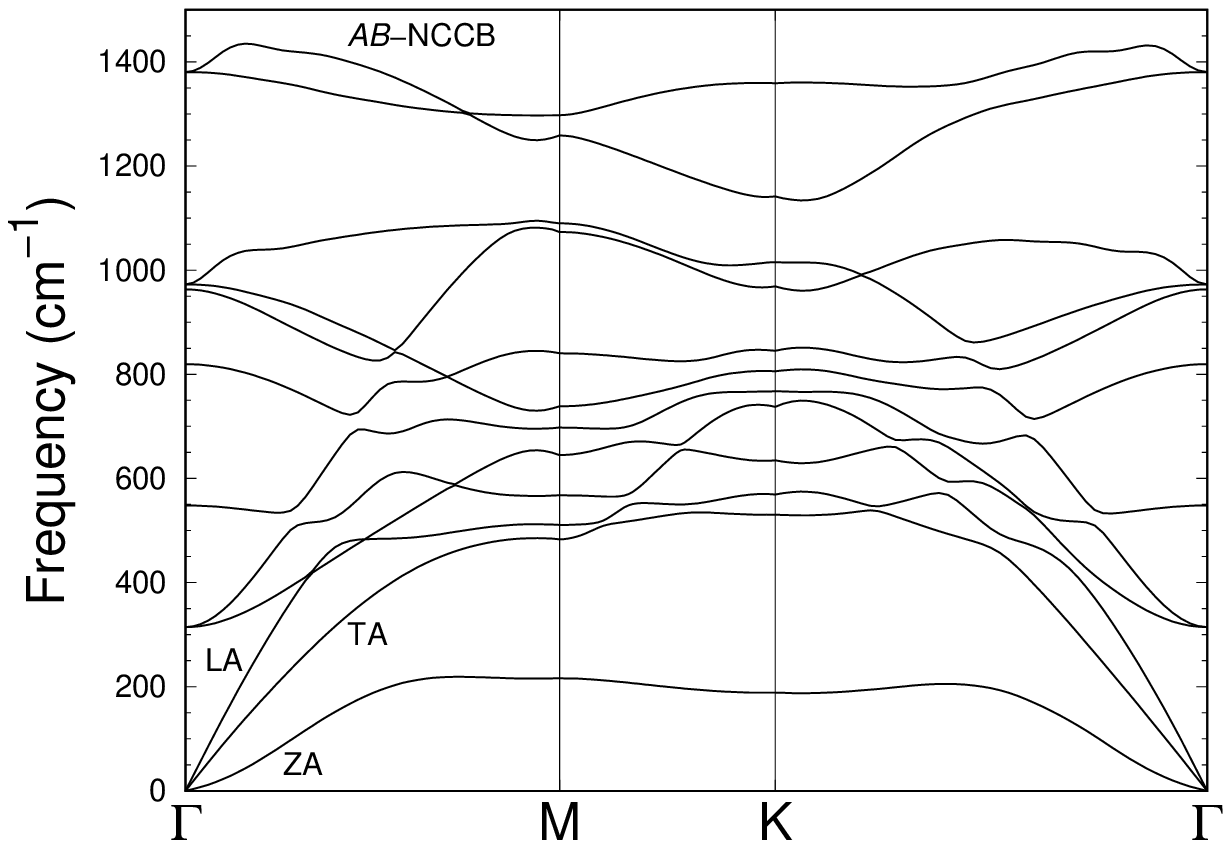}} 
 \subfloat[]{
 \includegraphics[scale = 0.45, trim={0cm 0.0cm 0cm 0.0cm}, clip]{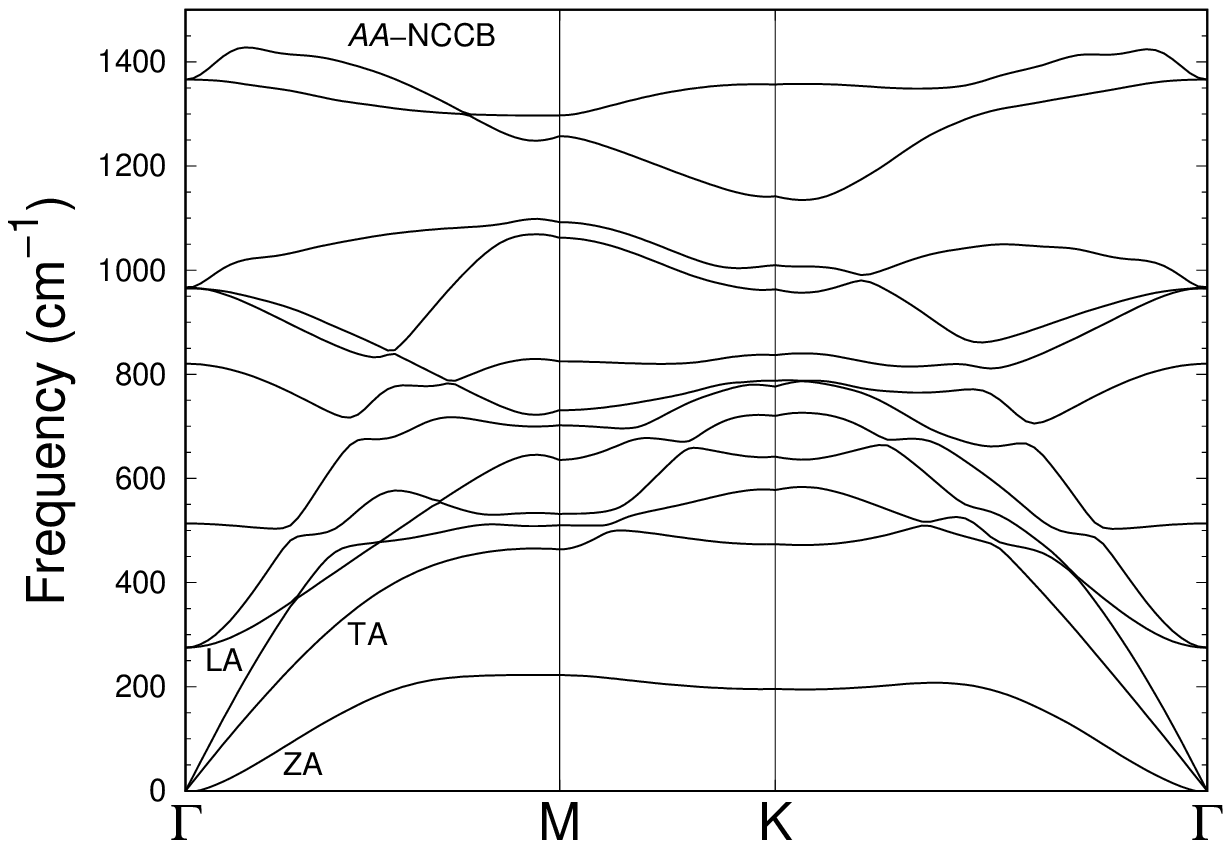}}
 \caption{Phonon dispersion of (a) $AB$-NCCN, (b) $AA$-NCCN, (c) $AB$-NCNC, (d) $AA$-NCNC, (e) $AB$-NCCB, and (f) $AA$-NCCB along the main high symmetry directions of the BZ of the hexagonal lattice.}
 \label{phonon_2}
 \end{figure}

The phonon dispersion curves of these stable structures are presented in figure \ref{phonon_2}, where the primitive cell structure contains 4 atoms and, hence, 12 phonon branches presenting only positives frequencies. which indicated structural stability. The in-plane acoustic modes, labeled as transverse (TA) and longitudinal (LA), exhibited linear variation at $q$ for $q \rightarrow 0$ in the $\Gamma$ point, while the out-of-plane ZA mode showed a quadratic dispersion at $q$ close to the $\Gamma$ point, which was similar to the theoretical and experimental results for graphene and graphite \cite{mounet2005first, mohr2007phonon}.

\textbf{Physical properties of bilayers}.
Table \ref{table1} presents the properties of the dynamically stable systems, as previously discussed: the NCCN, NCNC, and NCCB bilayers in the $AA$- and $AB$-stacking configurations. Among those structures, the $AB$-NCCN bilayer was the energetically most favorable one and, therefore, its energy of formation was taken as the reference value. Then, this configuration was followed by the $AA$-NCCN one, whose energy of formation was only 35 meV higher than that with the $AB$-stacking.

\begin{table}[htb]
\caption{Structural properties of NCCN, NCNC, and NCCB bilayers in $AA$- and $AB$-stackings: lattice parameter ($a$), inter-layer distances ($h$) and interatomic distances within the layers ($d$ and $d'$), and bond angles ($\alpha, \beta, \gamma$, and $\delta$), as defined in figure \ref{AA_AB}. $E_g$ is the bilayer band-gap and $\Delta E_f$ is the relative energy of formation, with respect to the $AB$-NCCN system. Distances, angles, and energies are given in {\AA}, degrees, and eV, respectively. Results of another theoretical investigation \cite{bondarchuk2017}, given in parenthesis, were obtained for a graphite-like bulk system by using the generalized  gradient  approximation (GGA-PBE) functional to describe the exchange-correlation term within the DFT, without considering the dispersive vdW interaction.}
\begin{center}
\begin{tabular}{cccccccc} \hline \hline
\multicolumn{1}{c}{stacking}&\multicolumn{3}{c}{$AA$} &\multicolumn{1}{c}{$\,\,\,\,\,$} &
\multicolumn{3}{c}{$AB$} \\
\multicolumn{1}{c}{struture}&\multicolumn{1}{c}{NCCN} & \multicolumn{1}{c}{NCNC} & \multicolumn{1}{c}{NCCB} &\multicolumn{1}{c}
{$\,\,\,\,\,\,\,\,$} &\multicolumn{1}{c}{NCCN} & \multicolumn{1}{c}{NCNC} & \multicolumn{1}{c}{NCCB} \\ \hline
   $a$          & 2.379  &  2.375 &  2.557       &  & 2.392(2.395)   &  2.394(2.394)     &  2.557   \\  
	 $h$          & 1.635  &  1.914 &  1.672     &  & 1.576          &  1.599            &  1.659   \\
	 $d$          & 1.465  &  1.436 &  1.564     &  & 1.471(1.475)   &  1.461(1.455)    &  1.559   \\
	 $d'$         & 1.465  &  1.462 &  1.495     &  & 1.471(1.475)   &  1.507(1.508)    &  1.494   \\
$\alpha$        & 108.6  &  111.5 &  109.6       &  & 108.8          &  110.0            &  110.2   \\
$\beta$         & 110.3  &  107.3 &  109.3       &  & 110.2(110.3)   &  108.9(108.2)    &  108.7    \\
$\gamma$        & 110.3  &  110.3 &  $\,\,99.0$  &  & 110.2(110.3)   &  113.3(113.5)    &  $\,\,98.8$ \\
$\delta$        & 108.6  &  108.6 &  117.6       &  & 108.8          &  113.6            &  117.7    \\
$E_g$           & 3.910  &  1.114 &  1.771       &  & 4.637(3.638)   &  2.299(1.948)    &  1.641    \\ 
$\Delta E_f$    & 0.035  &  2.218 &  0.687       &  & 0.000          &  2.346            &  0.486    \\ \hline \hline
\label{table1}
\end{tabular}
\end{center}
\end{table}

In the $AB$-NCCN bilayer, the C-N interatomic distances, $d$ and $d'$, were 1.471 {\AA}, close to the upper limit distance of the widely studied g-C$_3$N$_4$ \cite{zhu2018first}. The distance $h$ between layers, i.e. the C-C interatomic distance, was 1.576 {\AA}, very close to the value of 1.535 {\AA} of the C-C interatomic distance obtained in the diamond crystal, indicating that the $AB$-stacking of two h-CN layers did not give rise to a vdW system. In fact, the results indicated that the inter-layer bonding was primarily covalent. The N-C-N bond angles, $\alpha$ and $\delta$, were 108.8$^{\circ}$ and the N-C-C bond angles, $\beta$ and $\gamma$, were 110.2$^{\circ}$, being, respectively, 0.64\% smaller and larger than the C-C-C tetrahedral bond angle in the diamond crystal, indicating again the prevailing $sp^3$ character of those bonds. This differ substantially from a graphite-like $sp^2$ bonding, found in fullerenes, nanotubes, and graphene. All results were in good agreement with the ones obtained by another investigation for the graphite-like NCCN crystal \cite{bondarchuk2017}, as indicated in table \ref{table1}.

\begin{figure}[h!]
 \centering
 \subfloat[]{
 \includegraphics[scale = 0.4, trim={0cm 0.0cm 6.5cm 0.0cm}, clip]{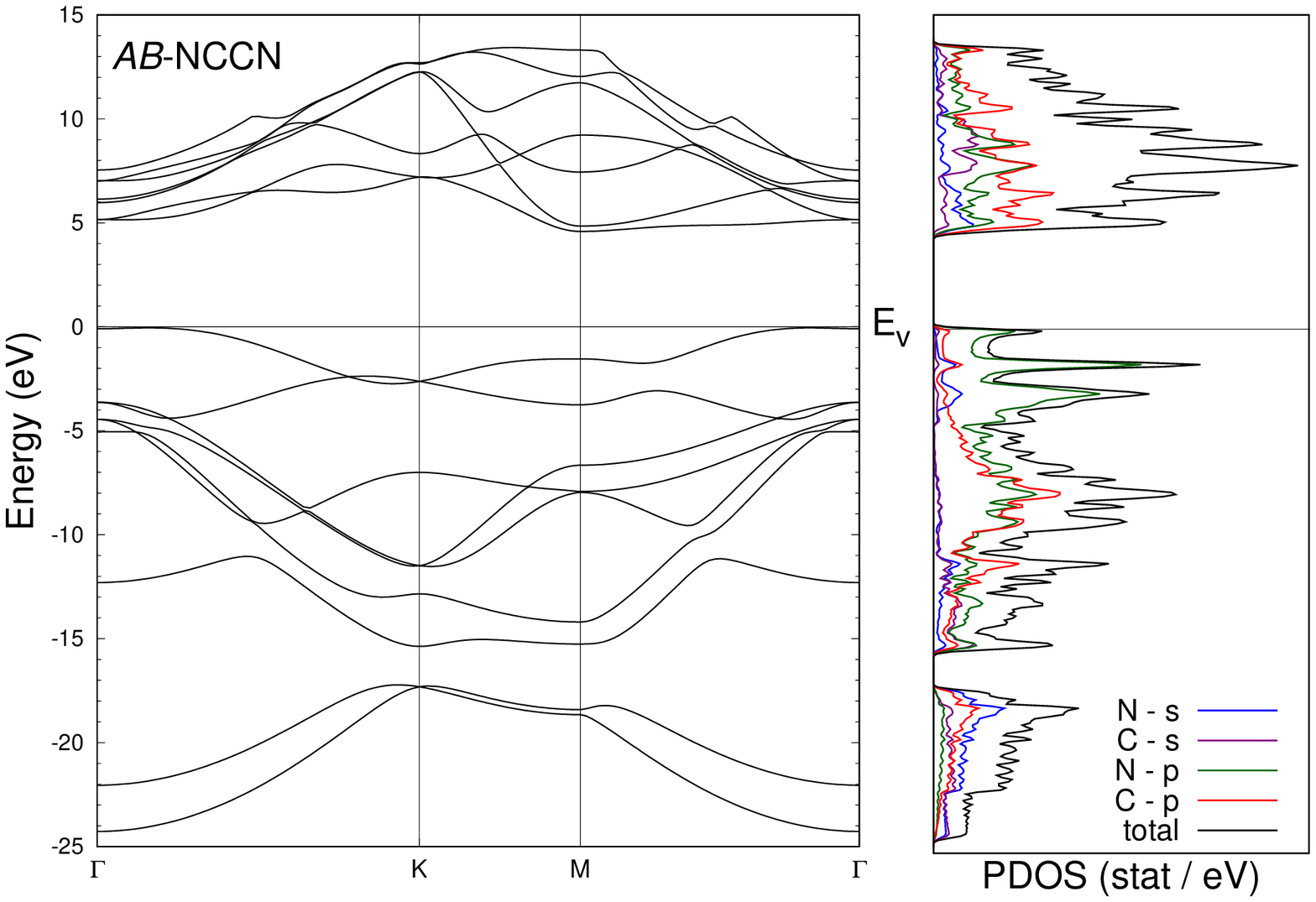}} 
 \subfloat[]{
 \includegraphics[scale = 0.4, trim={0cm 0.0cm 6.5cm 0.0cm}, clip]{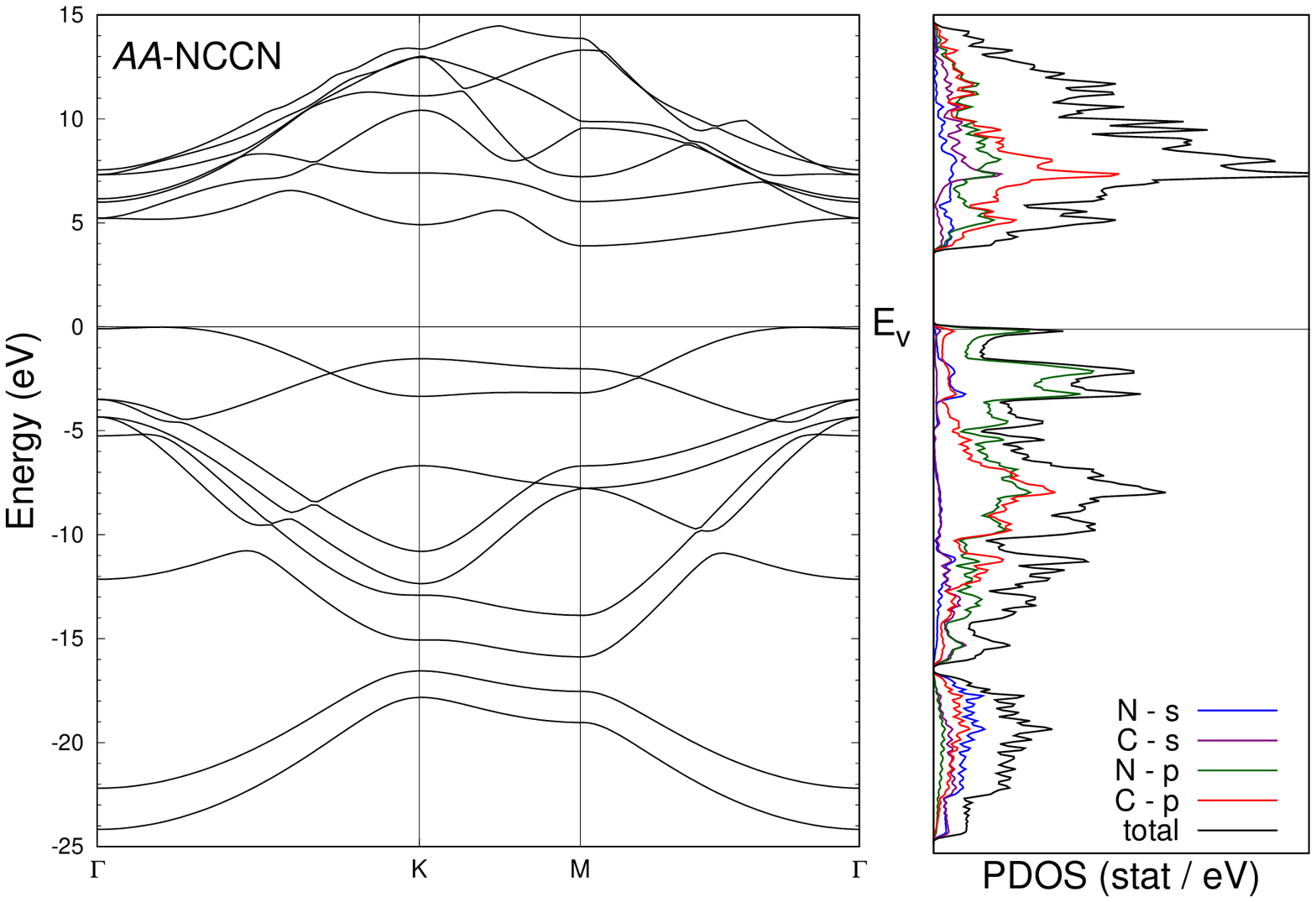}} \quad
 \subfloat[]{
 \includegraphics[scale = 0.4, trim={0cm 0.0cm 6.5cm 0.0cm}, clip]{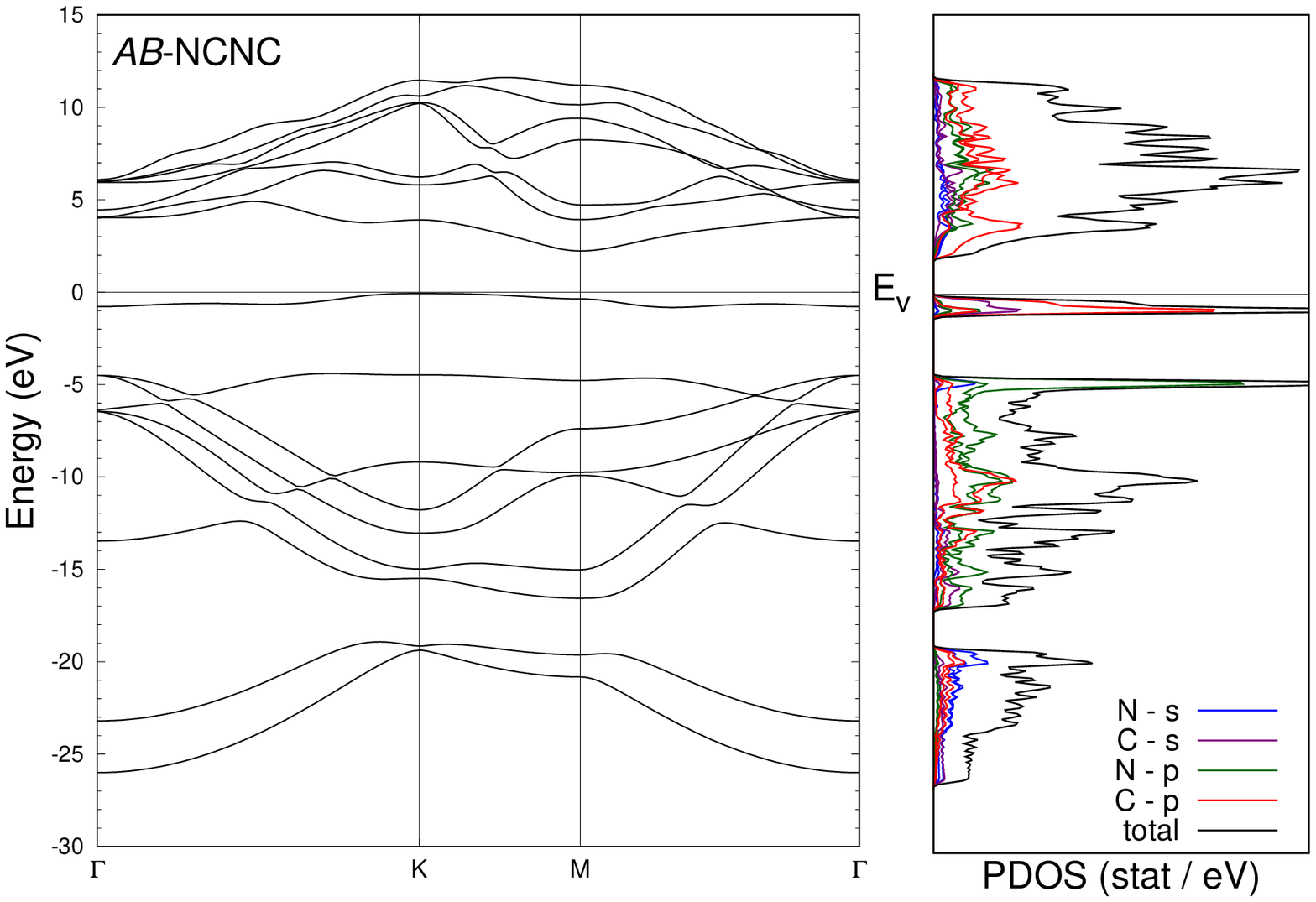}}
 \subfloat[]{
 \includegraphics[scale = 0.4, trim={0cm 0.0cm 6.5cm 0.0cm}, clip]{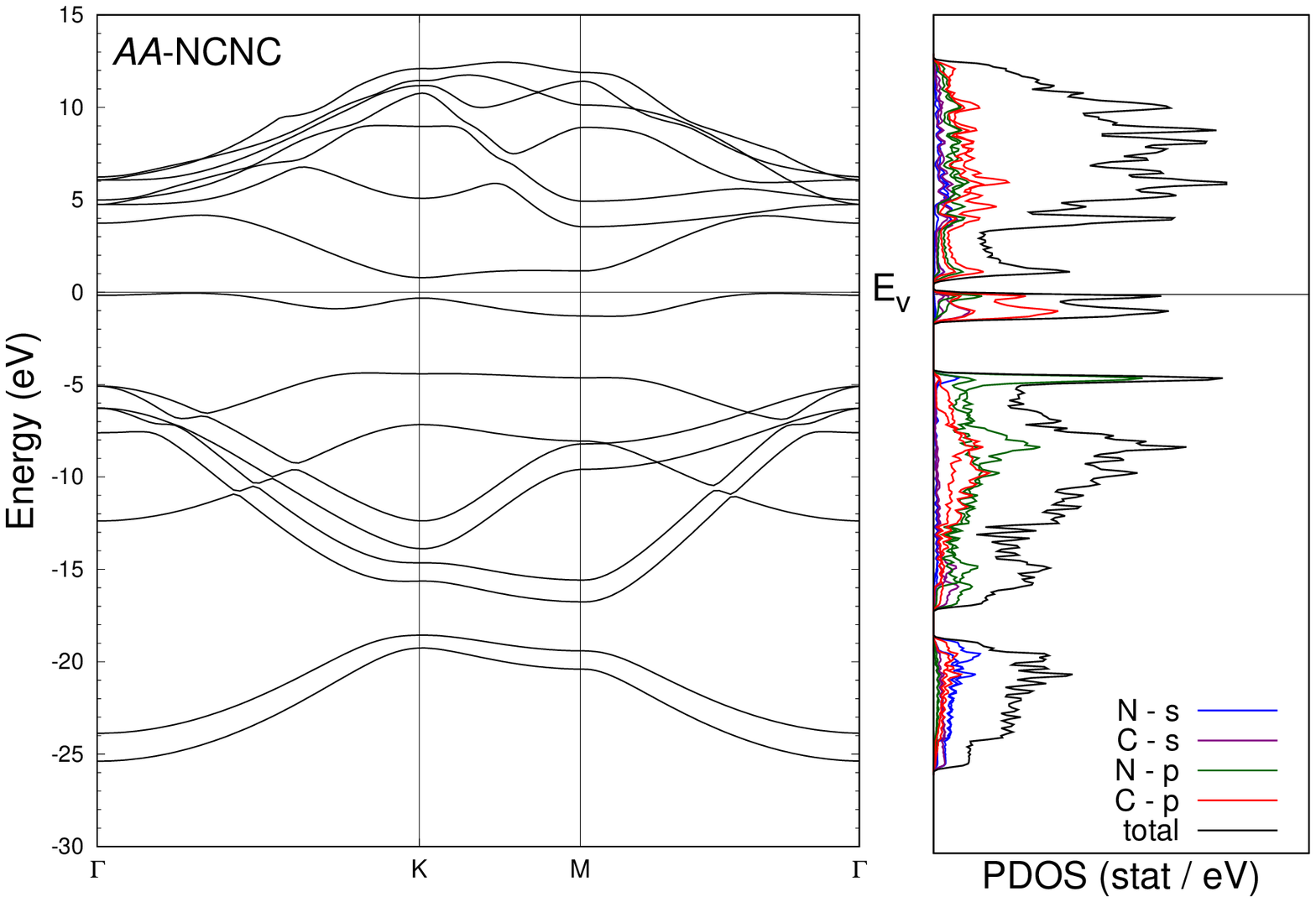}} \quad
 \subfloat[]{
 \includegraphics[scale = 0.4, trim={0cm 0.0cm 6.5cm 0.0cm}, clip]{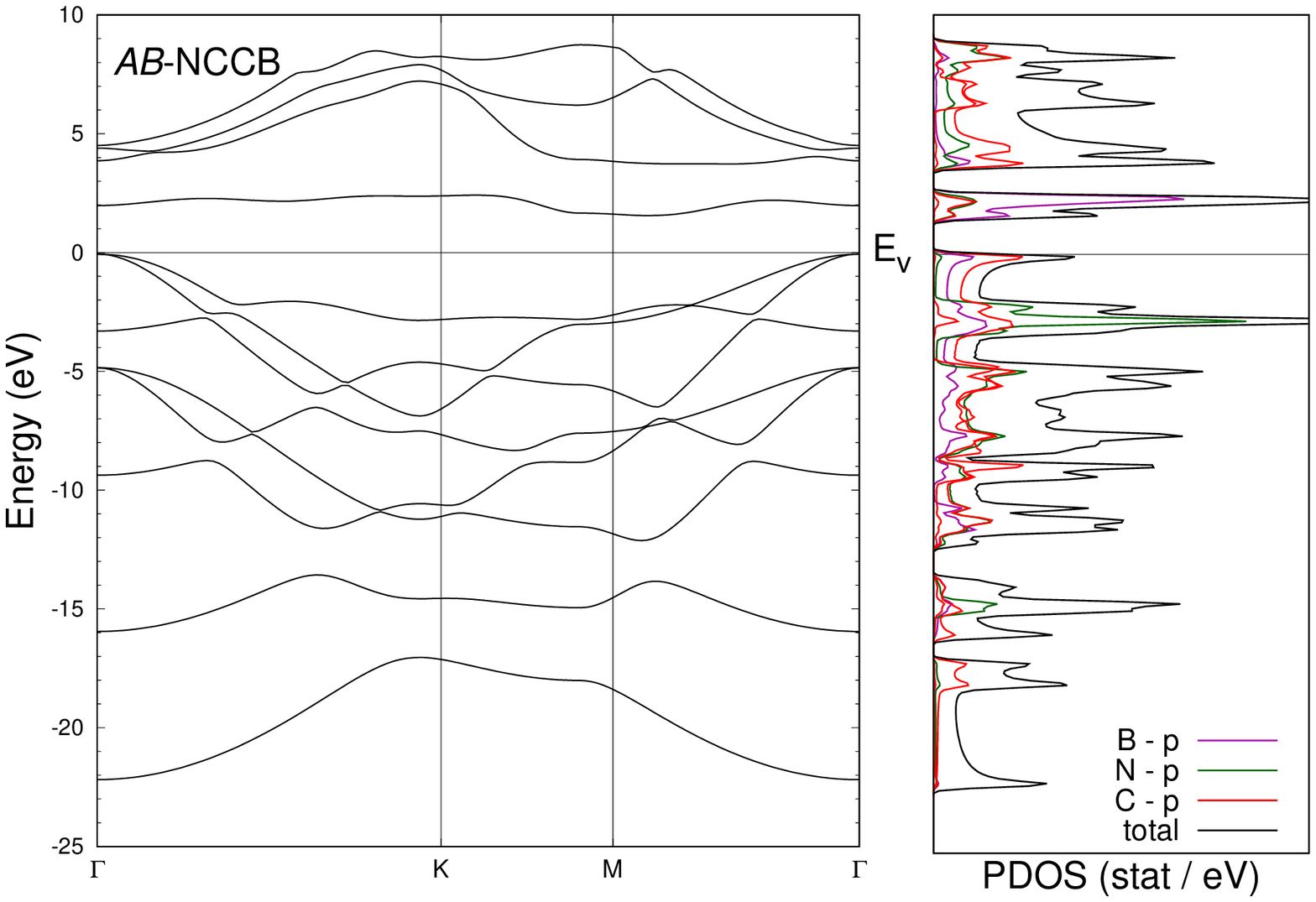}}
 \subfloat[]{
 \includegraphics[scale = 0.4, trim={0cm 0.0cm 6.5cm 0.0cm}, clip]{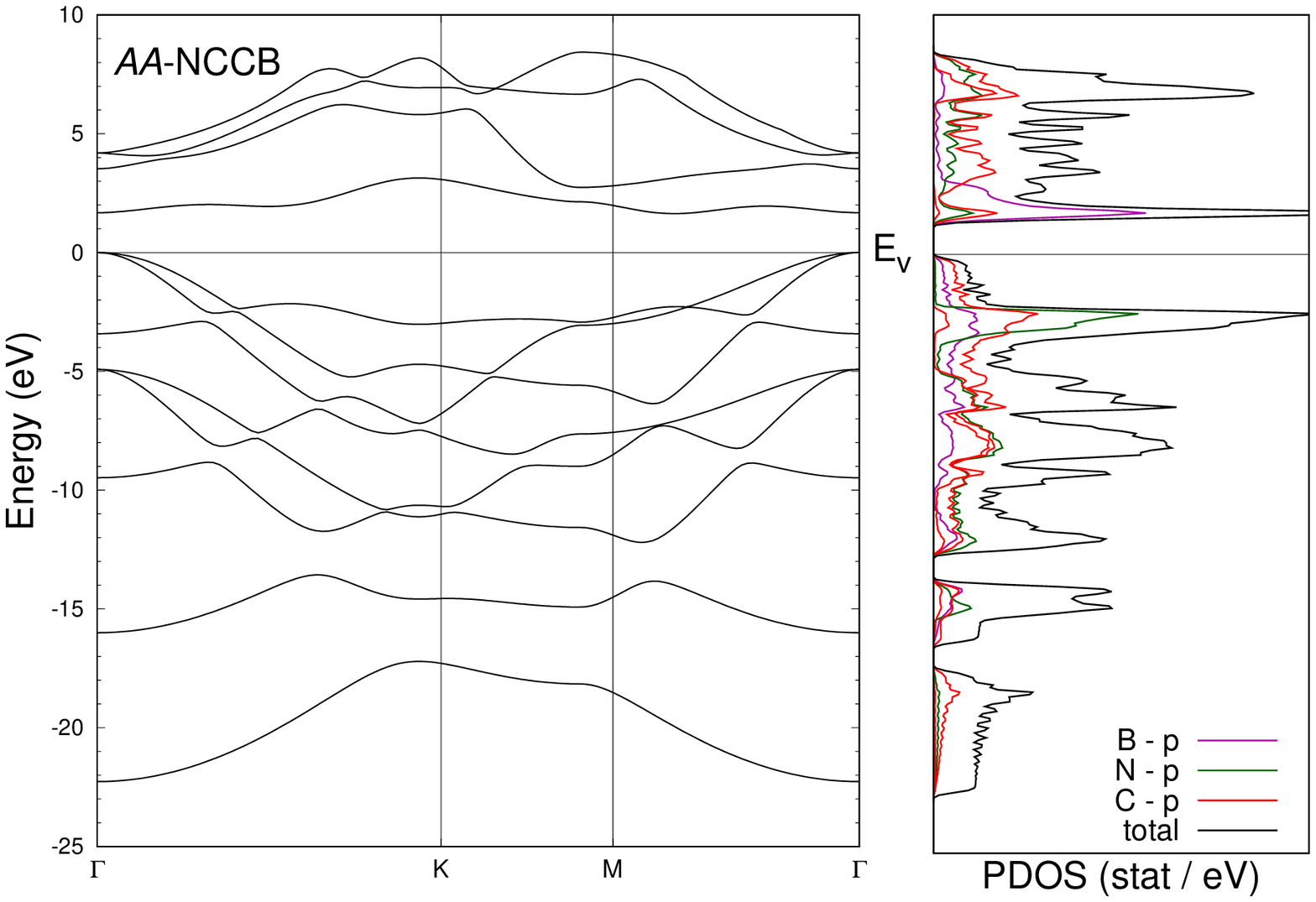}}
 \caption{Electronic band structures of (a) $AB$-NCCN, (b) $AA$-NCCN, (c) $AB$-NCNC, (d) $AA$-NCNC, (e) $AB$-NCCB, and (f) $AA$-NCCB configurations, along the main high-symmetry directions of the BZ. The figure also shows the total (black) and projected density of states on the $s$ orbitals of C (purple) and N (blue) atoms, and on the $p$ orbitals of C (red), N (green), and B (pink) atoms, in units of number of states/eV. E$_{\text{v}}$ represents the valence band top.}
 \label{pdos_2}
 \end{figure}

Furthermore, the $AB$-NCCN configuration presented an indirect electronic gap of 4.637 eV, with the highest occupied state (E$_{\text{v}}$) at the $\Gamma$-point and the lowest unoccupied state (E$_{\text{c}}$) at the M-point. This gap value was larger than the one of crystalline diamond of 4.356 eV, computed using the same methodology. The gap value for the $AB$-NCCN configuration was also larger than the value of 3.638 eV obtained by another investigation for the graphite-like NCCN crystal \cite{bondarchuk2017}. However, all those theoretical values should be considered as lower limits for the real gap since the DFT/vdW generally underestimates gap values. Figure \ref{pdos_2} (a) shows the $AB$-NCCN electronic band structure and total (DOS) and projected (PDOS) density of states on the C and N atomic orbitals. The valence band top had a prevailing $p$-N character with some contribution from the $s$-N, $p$-C, and $s$-C states, whereas the conduction band bottom had mainly contributions from the $p$-C related states with some contribution from the $p$-N states. Figure \ref{pdos_2} also shows the band structures of the $AA$-NCCN, $AB$-NCNC, $AA$-NCNC, $AB$-NCCB, and $AA$-NCCB configurations.

The $AA$-NCCN bilayer had the energy of formation just 35 meV over the $AB$-NCCN one and both structures had similar physical properties. For $AA$-NCCN, the C-N interatomic distances, $d$ and $d'$, were 1.465 {\AA} and the distance $h$ between layers was 1.635 {\AA}. For the bond angles, the in-plane ones ($\alpha$ and $\delta$) were 108.6$^{\circ}$, while both out-of-plane angles ($\beta$ and $\gamma$) were 110.3$^{\circ}$. The  $AA$-NCCN bilayer had the electronic band structure, shown in Fig. \ref{pdos_2} (b), close to that for $AB$-NCCN, however with a smaller gap of 3.910 eV.

The NCNC bilayers, with $AB$- or $AA$- stackings, were also dynamically stable. On the other hand, those bilayers presented energy of formation much higher than the one of the reference system ($AB$-NCCN). Therefore, despite being dynamically stable, they are likely inaccessible or considerably more difficult to grow in thermodynamic equilibrium conditions. In the NCNC structure, the C-N interatomic distances ($d$ and $d'$), the bond angles and the distance between layers ($h$) were strongly affected, when compared to the values of the NCCN structures, as presented in table \ref{table1}. For the $AB$-NCNC, the distance $h$ between layers was 1.599 {\AA}, slightly larger than the one in the $AB$-NCCN, $d$ and $d'$ were respectively 1.461 {\AA} and 1.507 {\AA}, while the in-plane bond angles, $\alpha$ and $\delta$, and the out-of-plane ones, $\beta$ and $\gamma$, were 110.0$^{\circ}$, 113.6$^{\circ}$, 108.9$^{\circ}$, and 113.3$^{\circ}$, respectively. These results were in good agreement with values obtained by another investigation in the literature for the graphite-like NCNC bulk \cite{bondarchuk2017}. The indirect electronic gap of this structure was 2.299 eV, which was larger than the value of 1.948 eV obtained by that investigation. Figures \ref{pdos_2} (c) and \ref{pdos_2} (d) show respectively the $AB$-NCNC and $AA$-NCNC electronic band structure and total (DOS) and projected (PDOS) density of states, on the C and N atomic orbitals. According to figure \ref{pdos_2} (c), the valence band top had a prevailing $p$-C character with some contribution from the $p$-N and $s$-C states, whereas the conduction band bottom had main contributions from the $p$-C states with some contribution from the $p$-N states. 

The $AA$-NCNC was energetically more stable than the $AB$-NCNC one by only 13 meV and their structural properties were slightly different. For this $AA$- stacking, the in-plane bond angles, $\alpha$ and $\delta$, and the out-of-plane ones, $\beta$ and $\gamma$, were 111.5$^{\circ}$, 108.6$^{\circ}$, 107.3$^{\circ}$, and 110.3$^{\circ}$, respectively. Moreover, the C-N interatomic distances, $d$ and $d'$, were respectively 1.436 {\AA} and 1.462 {\AA} and the distance $h$ between layers was 1.914 {\AA}, which was larger than the one in the $AB$-stacking. Additionally, the electronic gap, between the $\Gamma$- and K-points, was 1.114 eV, which was the smallest gap among the stable structures studied here. The atomic orbitals of the valence band top and conduction band bottom of the $AA$-NCNC had a similar composition of the $AB$-NCNC one. 

Replacing the N atoms by B ones in one side of an NCCN bilayer formed an NCCB bilayer. Although boron and nitrogen atoms have similar atomic sizes, the incorporation of the boron atoms led to important changes in the properties of the resulting structures. The C-N interatomic distances were 1.564 {\AA} and 1.559 {\AA} for $AA$- and $AB$-stacking, respectively, which were larger than in the NCCN and NCNC structures. The inter-layer distance $h$ (C-C interatomic distance) for both stackings, 1.672 {\AA} ($AA$-NCCB) and 1.659 {\AA} ($AB$-NCCB), were larger than the one in the diamond crystal. However, both of them still had a strong covalent inter-layer bond. The replacement of nitrogen (in NCCN) by boron, in order to build the NCCB structures, reduced the energy gap from 4.637 eV to 1.641 eV for the $AB$-stacking and from 3.910 eV to 1.771 eV for the $AA$-stacking.

The energy of formation of the $AA$-NCCB and $AB$-NCCB systems were respectively 0.687 eV and 0.486 eV higher than the reference value ($AB$-NCCN). Figures \ref{pdos_2} (e) and (f) present, respectively, the $AB$- and $AA$-NCCB electronic band structure and total (DOS) and projected (PDOS) density of states on the C, B, and N atomic orbitals. The atomic orbitals near the valence band top had major contributions from the $p$-C and $p$-B states with some contribution from the $p$-N states, whereas those near the conduction band bottom had mainly contributions from the $p$-B states with some contribution from the $p$-C and $p$-N states.

\begin{figure}[h!]
\centering
\subfloat[]{
\includegraphics[scale = 0.11, trim={2.5cm 0.0cm 4.5cm 0.0cm}, clip]{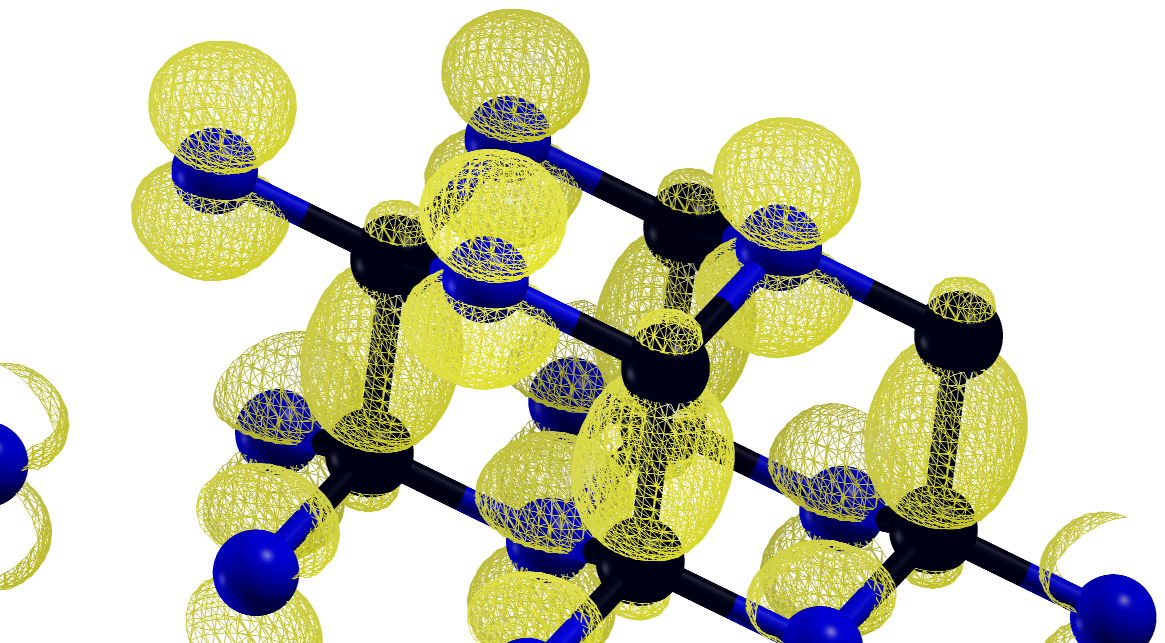}} 
\subfloat[]{
\includegraphics[scale = 0.09, trim={2cm 0.0cm 2cm 0.0cm}, clip]{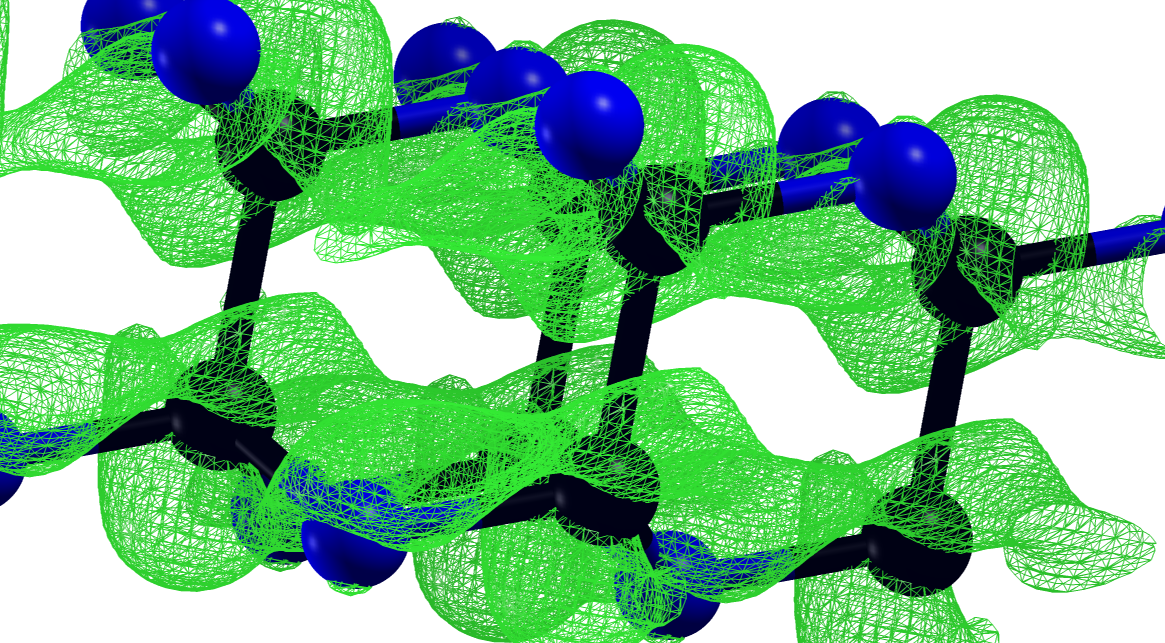}}  
\subfloat[]{
\includegraphics[scale = 0.11, trim={2cm 0.0cm 3cm 0.0cm}, clip]{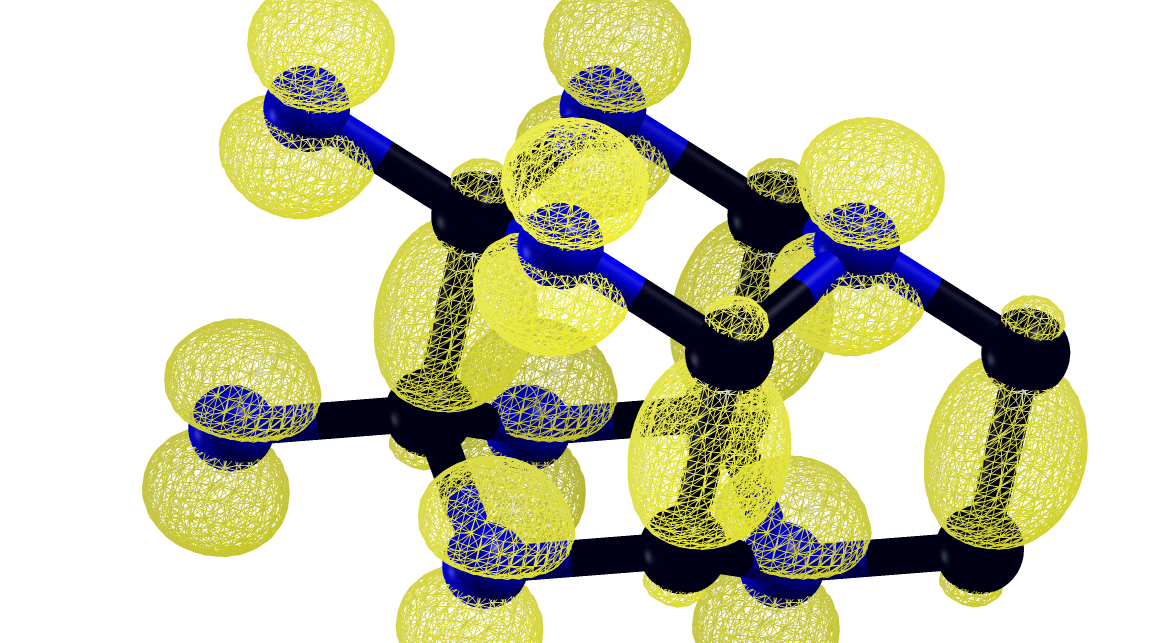}} 
\subfloat[]{
\includegraphics[scale = 0.11, trim={3.5cm 0.0cm 3.5cm 0.0cm}, clip]{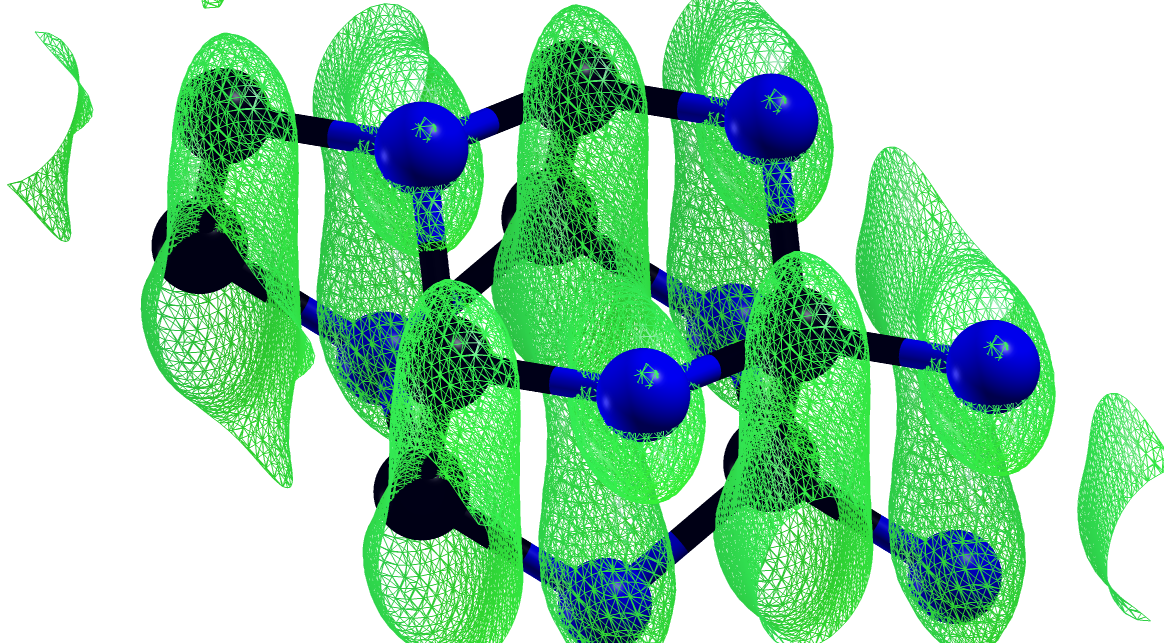}}  \\
\subfloat[]{
\includegraphics[scale = 0.10, trim={2.5cm 0.0cm 3cm 0.0cm}, clip]{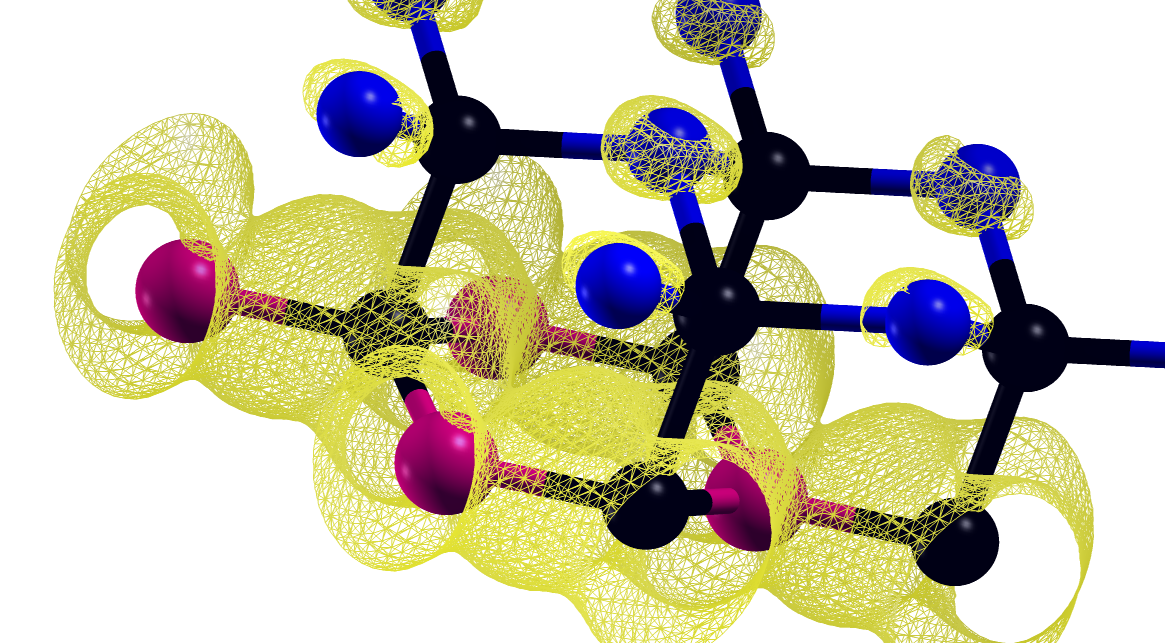}} 
\subfloat[]{
\includegraphics[scale = 0.09, trim={2.5cm 0.0cm 2cm 0cm}, clip]{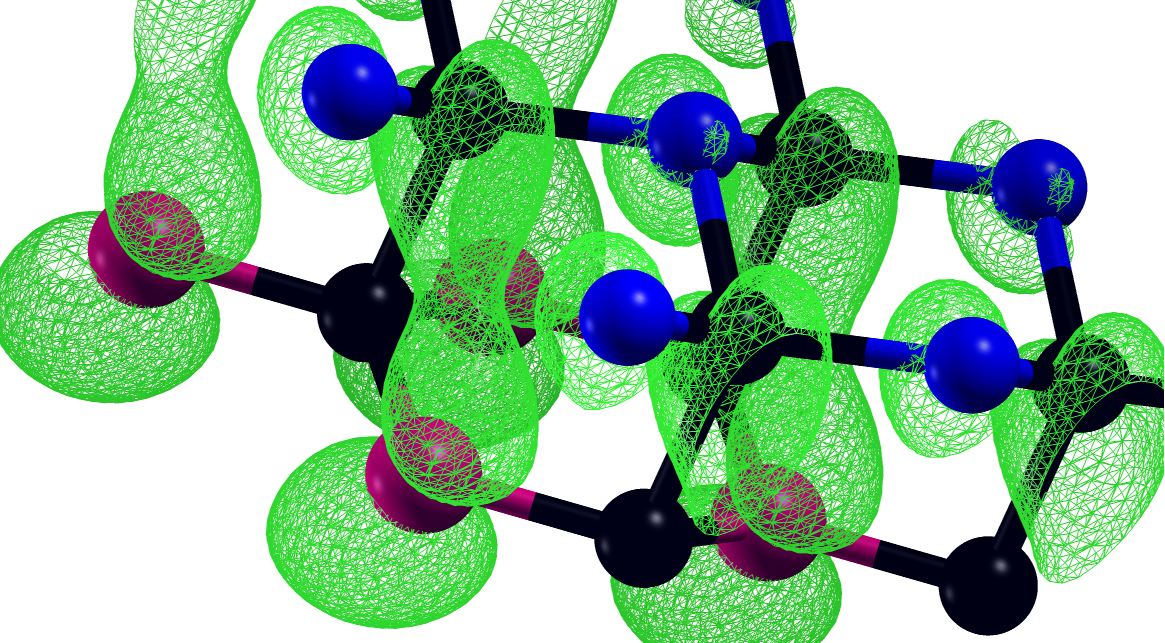}}  
\subfloat[]{
\includegraphics[scale = 0.11, trim={2.5cm 0.0cm 1.0cm 0.0cm}, clip]{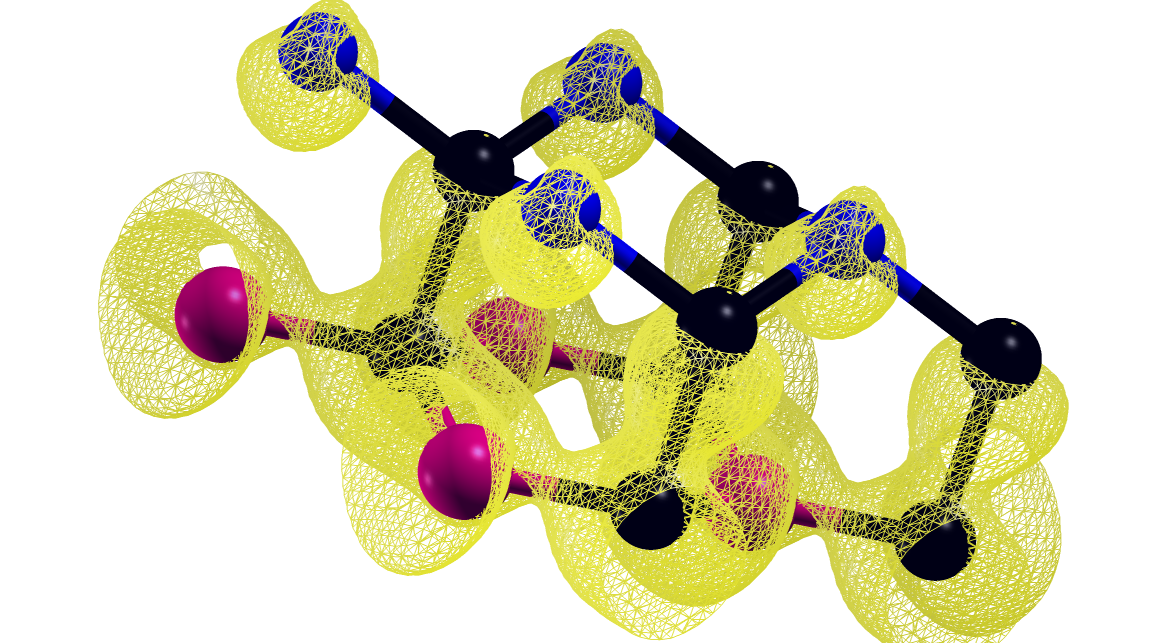}}
\subfloat[]{
\includegraphics[scale = 0.10, trim={2.5cm 0.0cm 2.5cm 0cm}, clip]{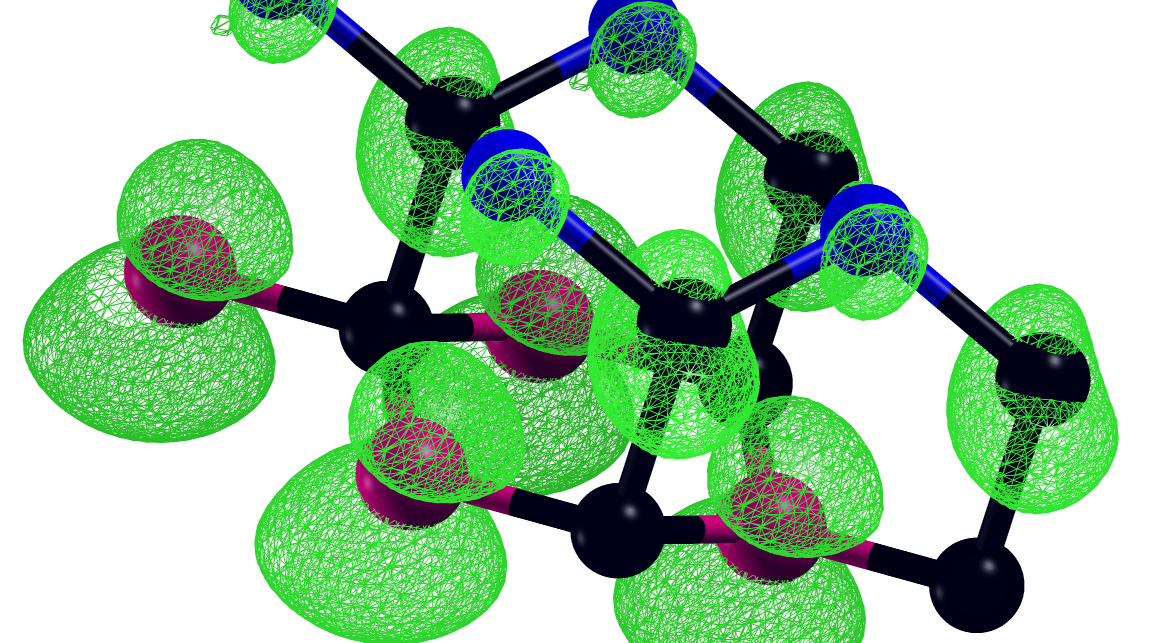}} \quad
\caption{Electronic charge density distributions of the NCCN and NCCB bilayers in the region around E$_{\text{v}}$ for (a) $AB$- and (c) $AA$-NCCN, and for (e) $AB$- and (g) $AA$-NCCB. The figure also shows the distributions in the region around E$_{\text{c}}$ for (b) $AB$- and (d) $AA$-NCCN, and for (f) $AB$- and (h) $AA$-NCCB.}
\label{Ev_Ec_AB_NCCN}
\end{figure}

Figure \ref{Ev_Ec_AB_NCCN} presents the probability density distributions on the region around E$_{\text{v}}$ and E$_{\text{c}}$. For $AA$- and $AB$-NCCN, the states near E$_{\text{v}}$ were associated with the C-C interatomic bonds plus the N-lone pairs (non-bonding), while the states around E$_{\text{c}}$ were associated with the C-C plus the C-N bonds, as shown in Fig. \ref{Ev_Ec_AB_NCCN} (a)-(d). Regarding $AA$- and $AB$-NCCB bilayers, as exhibited in Fig. \ref{Ev_Ec_AB_NCCN} (e)-(h), the density around E$_{\text{v}}$ was related mainly to the C-C and C-B interatomic bonds, and the one around E$_{\text{c}}$ was related to the C-N bonds with an antibonding character, distributed in the backbond of the N atoms. These probability density distributions provided complementary information for the analysis of the weight contributions of each $s$ and $p$ orbitals on band structures, DOS, and PDOS presented in Fig. \ref{pdos_2}.

\section{Discussion}
\textbf{2D Building blocks}.
Considering the stable structures presented previously, it is interesting to explore their potential applications, in the context of self-assembly growth. The requirements for self-assembly include a proper set of building blocks, a natural driving force between the building blocks for growth, and a final structure that is strong enough to be of practical use. As shown earlier, we found a set of stable bilayers that could be used as 2D building blocks and could provide strong bonding between different bilayers.

Although the NCNC bilayers had different atomic edges (associated to carbon or nitrogen atoms), which could help self-assembly growth, we discarded them as potential building blocks, since their energies of formation were much larger than the one of other bilayers. Therefore, the NCCN and NCCB, either in $AB$- or $AA$- stackings, represent potential building blocks. The NCCN bilayer should be considered as a primary building block candidate. It had the smallest energy of formation of all the structures studied here and carried reactive edges in the nitrogen-like surfaces. However, the ideal building block partner for an NCCN bilayer would be a BCCB bilayer, which was dynamically unstable and, therefore, unsuitable to serve as a building block. Even if a BCCB bilayer were dynamically stable, there would still be a challenge in placing an NCCN bilayer over a BCCB one due to a lattice mismatch between them. 

Without a partner bilayer, the NCCN bilayers do not seem appropriate to serve as a 2D building block for self-assembly. Considering it as a single building block, there would not exist a driving force between two neighboring NCCN bilayers stacked one over the other. However, the NCCN bilayers could still find its applications in other fields, such as serving as battery anodes to host lithium atoms or other large ions \cite{fu2014covalently, reddy2010synthesis}. In that case, the stacking of NCCN bilayers leads to very weak bonding between neighboring NCCN bilayers \cite{bondarchuk2017} and, hence, they could have available sites to host large ions. Nitrogen-doped graphene has also been widely investigated for several potential applications, including supercapacitors \cite{jeong2011nitrogen}, semiconducting devices \cite{wang2009n}, and hydrogen storage \cite{tang2013metal}. Moreover, although it has been reported  that a hexagonal two-dimensional carbon nitride structure should be unstable with N concentration exceeding 37.5\% \cite{shi2015}, the NCCN structures investigated here were remarkably stable and, therefore, they could serve as a guide to the discovery of new carbon-based materials with high N concentration.  

Therefore, an NCCB bilayer seemed to be the most promising 2D building block, in terms of a large reactivity and a small energy of formation. An NCCB bilayer could be stacked over another NCCB one, a stacking that could benefit from the strong nitrogen-boron interactions. Accordingly, the NCCB bilayer caried all the requirements to be used as a building block in self-assembly growth processes.

\textbf{Functionalized graphene-like bilayer crystals}.
In order to explore the possibility of an NCCB serving as a 2D building block, we investigated the properties of 3D crystals made by stacking those building blocks.

\begin{figure}[hbt]
\centering
\subfloat[]{
\includegraphics[scale = 0.21, trim={2.5cm 0.0cm 4.5cm 0.0cm}, clip]{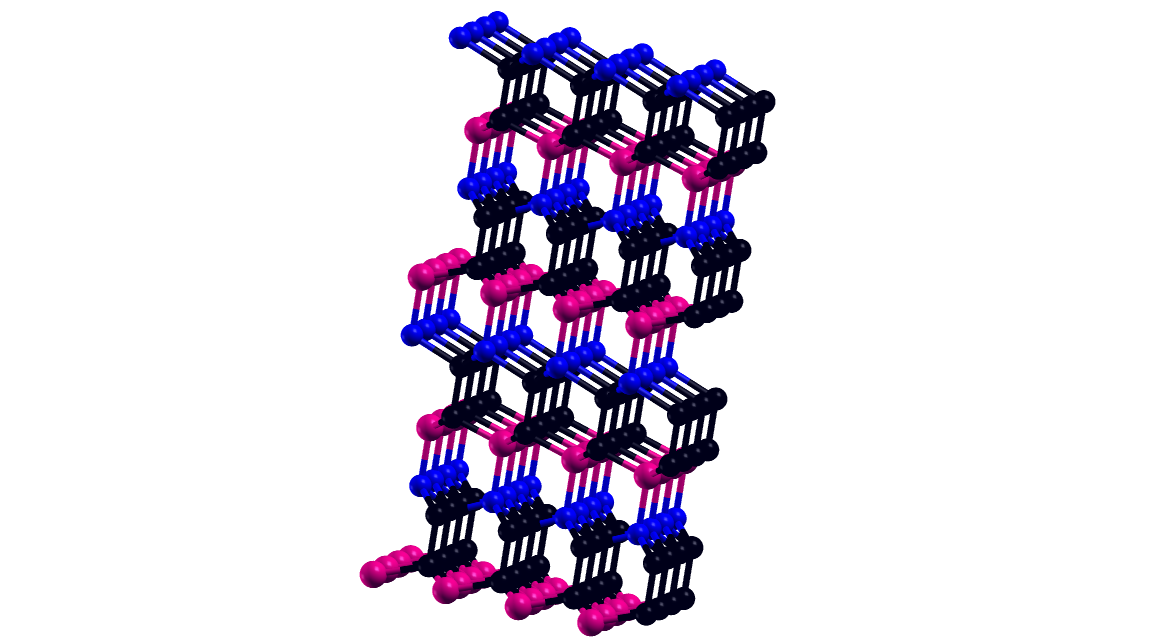}}
\subfloat[]{
\includegraphics[scale = 0.21, trim={2.5cm 0.0cm 4.5cm 0.0cm}, clip]{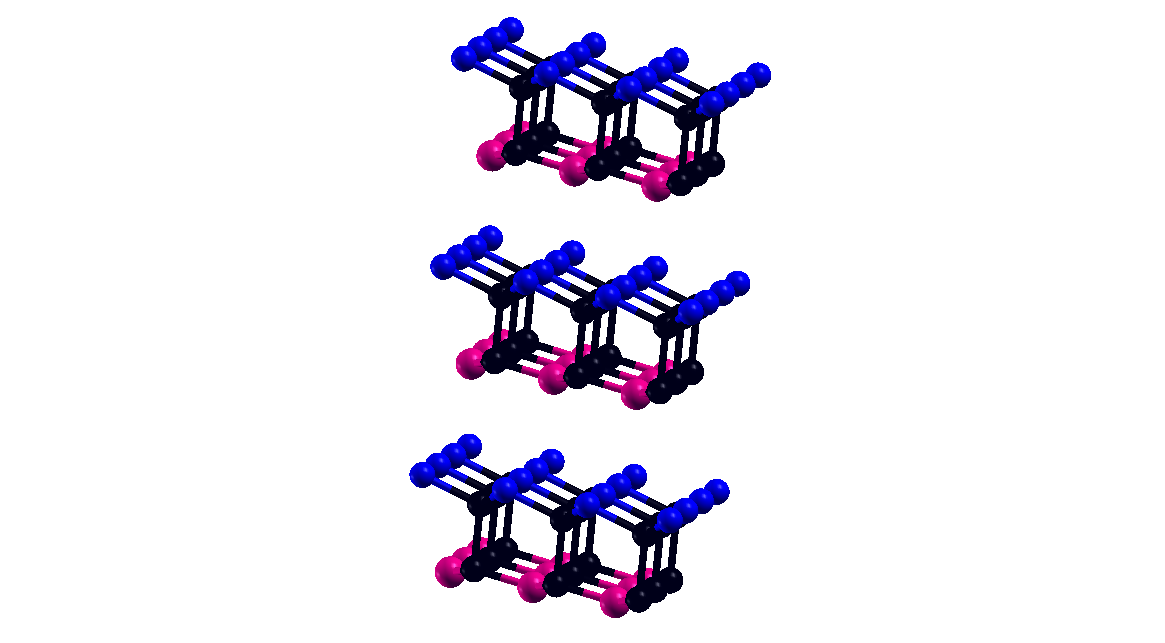}}\\
\subfloat[]{
\includegraphics[scale = 0.35, trim={0cm 0.0cm 6.5cm 0.0cm}, clip]{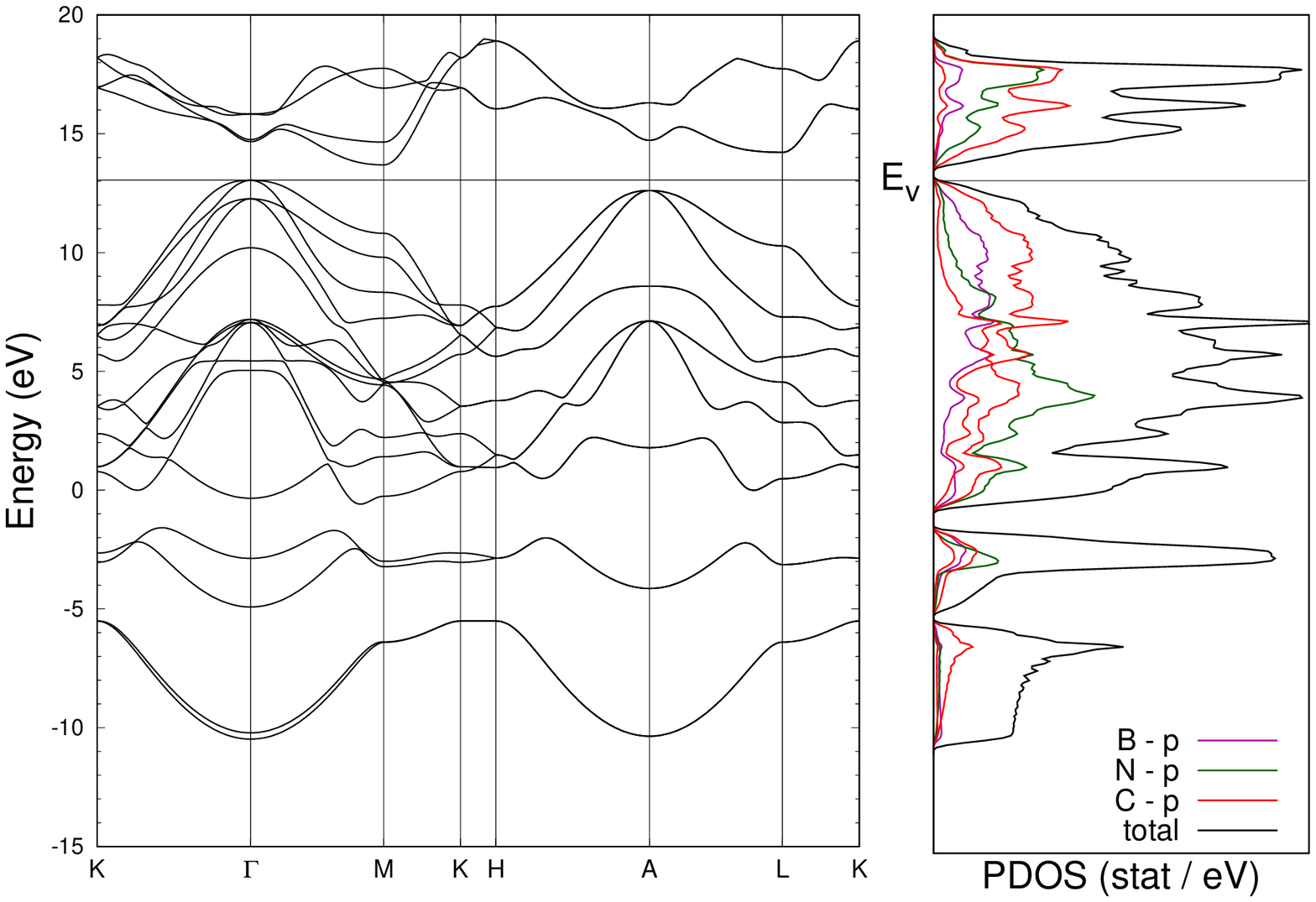}}
\subfloat[]{
\includegraphics[scale = 0.35, trim={0cm 0.0cm 6.5cm 0.0cm}, clip]{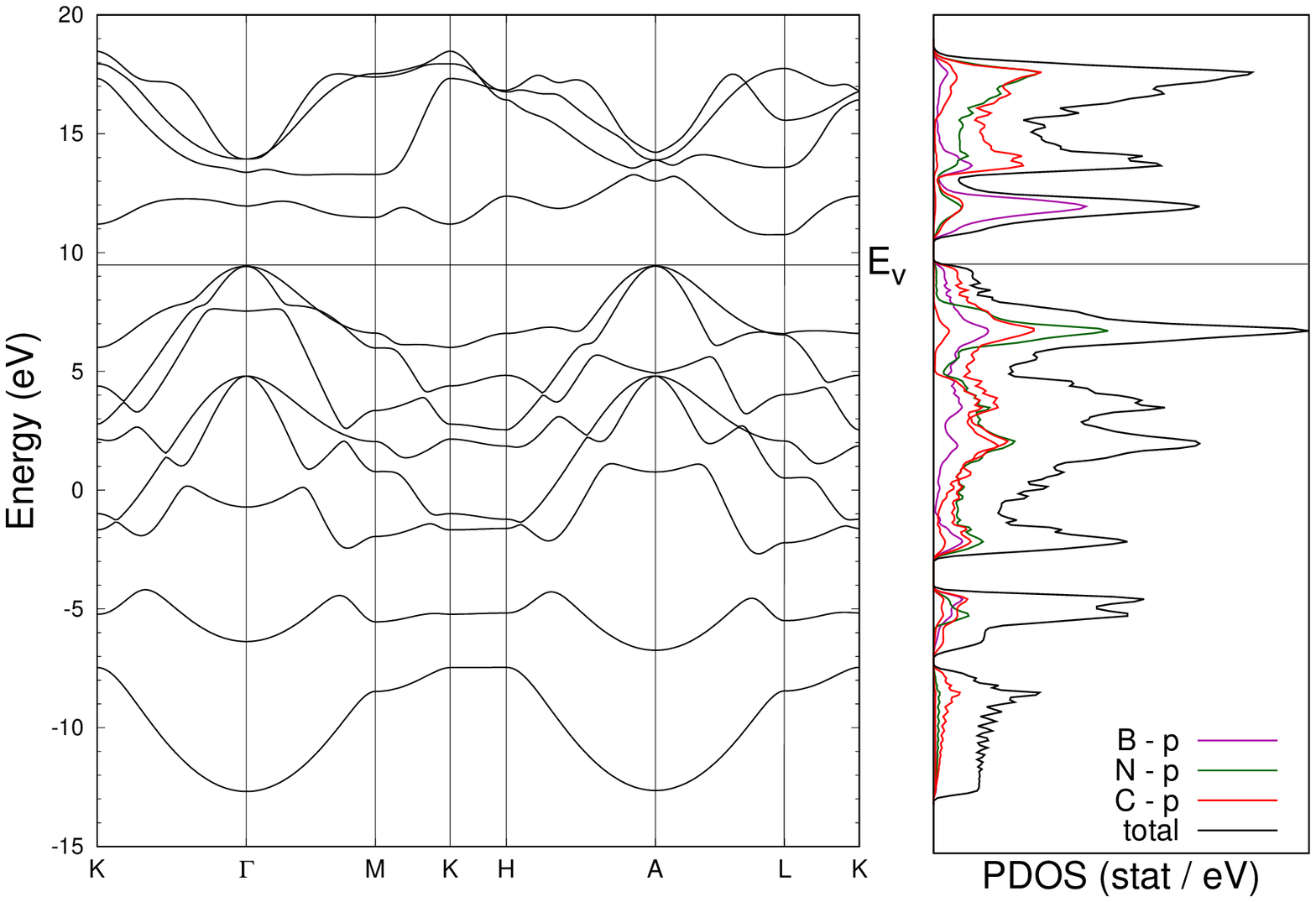}}
\caption{Schematic representation of (a) the NCCB covalent crystal and (b) the NCCB graphite-like one, where the black, pink, and blue spheres represent carbon, boron, and nitrogen atoms, respectively. Electronic band structures of the (c) covalent-like and (d) graphite-like crystal configurations, along the main high-symmetry directions of the BZ. The figure also presents the total (black) and projected density of states on the $p$ orbitals of C (red), N (green), and B (pink) atoms, in units of number of states/eV. E$_{\text{v}}$ represents the valence band top.}
\label{crystal_NCCB_bb}
\end{figure}

The $AB$-NCCB bilayer was considered as the unitary block to construct two different crystalline structures in a hexagonal Bravais lattice. The first one was built by stacking two bilayers and the optimized lattice parameters were $a =$ 2.590 {\AA} and $c =$ 8.417 {\AA}, where the boron atoms from one NCCB bilayer were located exactly above the nitrogen atoms from the adjacent NCCB bilayer, as shown in figure \ref{crystal_NCCB_bb} (a). The C-C intra-bilayer distance ($h$) was 1.439 {\AA}, which was smaller than the one in the diamond crystal but close to the in-plane graphite and graphene bond lengths, indicating a covalent bond between those carbon atoms. Additionally, the B-N inter-bilayers distance of 1.645 {\AA} was larger, for example, than the B-N length found in h-BN (1.45 {\AA}) \cite{wang2017graphene}, in the superhard cubic boron-carbonitride (1.575 {\AA}), and in hexagonal BC$_2$N (1.565 {\AA}) crystals \cite{sun2001structural, liu2018hexagonal}. However, this distance was similar to the B-N dative bonding found in several other structures \cite{brinck1993computational,Garcia}. The C-C covalent bond and B-N dative one could guarantee stability and rigidity of the resulting crystalline structure. 

Moreover, the self-assembly could be facilitated by the ionic-like driving force between N and B atoms from neighboring building blocks, since N has a donor character while B has an acceptor one. The values of the binding energy ($E_b$ = 0.16 eV/atom) and the bulk modulus of the crystal ($K_0$ = 328.9 GPa), which was obtained by fitting the third-order finite strain Birch-Murnaghan equation of state to the energy vs. volume relation \cite{birch1947finite}, indicated that this structure presented characteristics to be considered a hard material with covalent bonding \cite{grimsditch1994elastic}. The crystal presented a small indirect electronic gap of 0.65 eV, where the highest occupied state (E$_{\text{v}}$) was at the $\Gamma$-point and the lowest unoccupied state (E$_{\text{c}}$) was at the M-point, as shown in figure \ref{crystal_NCCB_bb} (c). The total (DOS) and projected (PDOS) density of states on the C, B, and N atomic orbitals are shown in figure \ref{crystal_NCCB_bb} (c) as well, where the valence band top had major contributions from the $p$-C and $p$-N states (C-N bond), whereas the conduction band bottom had mainly contributions from the $p$-C and $p$-B states (C-B bond).  

The second crystal was built in a graphite-like structure, with the optimized lattice parameters $a =$ 2.563 {\AA} and $c =$ 5.090 {\AA}. In this structure, the B and N atoms were located above/below the hexagonal hollow site of the adjacent bilayer, as displayed in figure \ref{crystal_NCCB_bb} (b). The C-C intra-bilayer distance was 1.587 {\AA}, close to that of the diamond crystal and of the same order to that in graphite and graphene, indicating a covalent bond between the carbon atoms.

Additionally, the inter-bilayer distance was 2.685 {\AA} and the bulk modulus 79.3 GPa. The value of 70 meV/atom obtained for the average binding energy was close to that found for graphite. These values suggest that this crystal could be classified as a layered material since it presented strong and covalent intra-bilayer bonds, but weak (van der Waals-like) inter-bilayers bonds. Furthermore, the crystal exhibited an indirect gap of 1.246 eV, in which the E$_{\text{v}}$ was at the A-point and the E$_{\text{c}}$ was at the L-point, as shown in figure \ref{crystal_NCCB_bb} (d). The figure presents the DOS and PDOS on the C, B, and N atomic orbitals as well, where the valence band top had major contributions from the $p$-C and $p$-B states, whereas the conduction band bottom had mainly contributions from the $p$-B states with some contribution from the $p$-C and $p$-N states.

{\section{Conclusions}}

In summary, we explored the properties of carbon-related monolayers and bilayers, functionalized with nitrogen and/or boron atoms. Those structures were investigated by state-of-the-art first-principles calculations, allowing to obtain information on stability, and structural and electronic properties. We found no dynamically stable functionalized carbon-related monolayers, such as h-CN or h-CB. Then, we explored the properties of graphene-related bilayers and found a set of dynamically stable structures formed by stacking N- and/or B- doped graphene-like monolayers. Particularly, several bilayers (labeled as NCCN, NCNC, and NCCB configurations) in two different stackings were stable. 

Of all the stable bilayers, the NCCN structure presented the smallest energy of formation, which would favor its growth, when compared to other functionalized bilayers. However, this NCCN structure does not seem an appropriate 2D building block, since there is no driving force for self-assembly. Moreover, the resulting stacking of NCCN layers does not lead to a tight macroscopic crystal. Still, this NCCN bilayer carries interesting physical properties to serve as battery anode to store lithium or other large ions. Moreover, the NCCN stability with N concentration of 50\% was a remarkable result, since a previous investigation indicated the upper limit for N concentration at 37.5\%, hence, these NCCN structures could lead to the exploration of new carbon nitride materials presenting a 1:1 stoichiometry.

Finally, the bilayers formed by nitrogen, carbon, and boron atoms, labeled as NCCB, were dynamically stable, with small energy of formation. They carried properties that make them very suitable to serve as 2D building blocks, with a strong ionic-like driving force between different bilayers. Therefore, the multiple stacking of NCCB bilayers should be very stable through self-organization, which is essential for any relevant technological application. This was confirmed by exploring the properties of the 3D crystals constructed by NCCB bilayers, indicating a configuration that could guarantee stability and rigidity, like the C-C covalent bonding and the B-N dative one. The resulting covalent 3D crystal presented a large bulk modulus and could be considered a hard material, providing another route in the search of superhard materials. On the other hand, the B-N dative bonding in the graphite-like crystal presented a weak van der Waals-like inter-bilayer interaction and a strong intra-bilayer covalent bonding, characteristics that lead to classify it as a layered material, such as graphite. Although both materials had the same building blocks as a fundamental cell, these results showed that different relative twist between bilayers changed radically the properties of the resulting crystals.

This investigation revealed that, while monolayers appear to be unsuitable to be used as 2D building blocks, bilayers are the primary candidates as building blocks. This indicates the possibility of exploring even other exotic 3D structures made of sandwiched layers with different atomic species or even with larger multilayer building blocks, such as trilayers.

\begin{acknowledgement}

This investigation was partially supported by Brazilian agencies CNPq and CAPES. We acknowledge resources from the Blue Gene/Q supercomputer supported by the High Performance Computing Center (Universidade de S\~ao Paulo - Rice University agreement).

\end{acknowledgement}

\providecommand{\latin}[1]{#1}
\makeatletter
\providecommand{\doi}
  {\begingroup\let\do\@makeother\dospecials
  \catcode`\{=1 \catcode`\}=2 \doi@aux}
\providecommand{\doi@aux}[1]{\endgroup\texttt{#1}}
\makeatother
\providecommand*\mcitethebibliography{\thebibliography}
\csname @ifundefined\endcsname{endmcitethebibliography}
  {\let\endmcitethebibliography\endthebibliography}{}

\end{document}